\documentclass[conference]{IEEEtran}
\IEEEoverridecommandlockouts
\usepackage{cite}
\usepackage{amsmath,amssymb,amsfonts}
\usepackage{algorithmic}

\usepackage{textcomp}
\usepackage[table]{xcolor}
\usepackage{enumitem}
\usepackage{booktabs} 
\usepackage{stfloats}
\usepackage{graphicx} 
\usepackage{subcaption} 
\usepackage{multirow} 

\usepackage{stmaryrd}

\usepackage{tikz}
\usepackage[ruled,vlined]{algorithm2e}

\usepackage{url}
\usepackage[T1]{fontenc}

\definecolor{green}{RGB}{0,100,0} 
\definecolor{Color_Hourglass}{RGB}{157,195,230} 
\definecolor{Color_CPA}{RGB}{146,208,80}        
\definecolor{Color_SSCPA}{RGB}{244,145,148}     
\definecolor{Color_Ladder}{RGB}{197,172,217}    
\definecolor{Color_Nested}{RGB}{244,177,131}    
\definecolor{Color_HOBA}{RGB}{254,108,108}      
\definecolor{Color_Thermos}{RGB}{105,141,207}   
\definecolor{Color_HHA}{RGB}{255,192,0}         
\definecolor{Color_Comm}{RGB}{255,255,255}      
\definecolor{arrowyellow}{RGB}{255,212,75} 

\definecolor{color_U}{RGB}{202, 232, 168} 
\definecolor{color_V}{RGB}{252, 235, 221} 

\definecolor{color_cpd}{RGB}{197, 123, 165} 

\definecolor{my_ring_purple}{RGB}{162,76,130}   
\definecolor{my_node_pink}{RGB}{233,166,189}    
\definecolor{my_node_green}{RGB}{169,209,142}   
\definecolor{my_core_gray}{RGB}{135,135,135}    

\definecolor{my_node_pink}{RGB}{234, 169, 188}    
\definecolor{my_node_green}{RGB}{166, 206, 148}   

\definecolor{lightblue}{RGB}{236,240,250} 
\definecolor{revisiondarkgreen}{RGB}{0,0,0} 

\newcommand{\CirclePink}{%
    \tikz[baseline=-0.6ex]{%
        \fill[my_node_pink] (0,0) circle (0.8ex);%
    }%
}

\newcommand{\CircleGreen}{%
    \tikz[baseline=-0.6ex]{%
        \fill[my_node_green] (0,0) circle (0.8ex);%
    }%
}

\newcommand{\YellowArrow}{%
\tikz[baseline=-0.5ex]{
   \draw[->, >=stealth, line width=1.5pt, color=arrowyellow] (0, 0) -- (2.5ex, 0);%
}%
}

\newcommand{\CircleU}{%
    \tikz[baseline=-0.6ex]{%
        \fill[color_U] (0,0) circle (0.8ex);%
    }%
}

\newcommand{\CircleV}{%
    \tikz[baseline=-0.6ex]{%
        \fill[color_V] (0,0) circle (0.8ex);%
    }%
}

\newcommand{\SymbolCPD}{%
    \tikz[baseline=-0.7ex, scale=1]{%
        \tikzstyle{s} = [color=color_cpd, line width=1.1pt, line cap=round];
        
        \draw[s] (0,0) circle (0.9ex);
        
        \draw[s] (155:0.9ex) -- (335:0.9ex);
        
        \draw[s] (135:0.9ex) -- (135:1.6ex); 
        \draw[s] (45:0.9ex)  -- (45:1.6ex);  
        \draw[s] (225:0.9ex) -- (225:1.6ex); 
        \draw[s] (315:0.9ex) -- (315:1.6ex); 
    }%
}

\newcommand{\SymbolG}{%
    \tikz[baseline=-0.5ex, scale=0.5]{
        
        \tikzstyle{core} = [draw=my_core_gray, line width=0.8pt];
        \tikzstyle{leg}  = [draw=my_core_gray, line width=0.8pt];
        
        \draw[color=my_ring_purple, line width=0.8pt] (0,0) circle (2.4ex);

        \draw[leg] (0,0) -- (140:1.4ex);
        \draw[leg] (0,0) -- (40:1.4ex);
        \draw[leg] (0,0) -- (270:1.4ex);

        \fill[my_node_pink] (140:1.4ex) circle (0.7ex);
        \fill[my_node_pink] (40:1.4ex)  circle (0.7ex);
        \fill[my_node_green] (270:1.4ex) circle (0.7ex);

        \draw[core, fill=white] (0,0) circle (0.7ex);
        
        \draw[core] (160:0.7ex) -- (340:0.7ex);
    }%
}

\def\BibTeX{{\rm B\kern-.05em{\sc i\kern-.025em b}\kern-.08em
    T\kern-.1667em\lower.7ex\hbox{E}\kern-.125emX}}
\begin{document}

\title{Sparse Array-Inspired Frame Structure Design for 5G NR ISAC
}


\author{Haisu Wu, Hong Ren,~\IEEEmembership{Member,~IEEE}, Cunhua Pan,~\IEEEmembership{Senior Member,~IEEE}, Yuqing Chen,\\ Ruizhe Wang, Jun Tang, Dongming Wang,~\IEEEmembership{Member,~IEEE}, and Jiangzhou Wang,~\IEEEmembership{Fellow,~IEEE}
\thanks{Haisu Wu, Hong Ren, Cunhua Pan, Ruizhe Wang, Jun Tang, Dongming Wang and Jiangzhou Wang are with National Mobile Communications Research Laboratory, Southeast University, Nanjing, China (e-mail: wuhaisu, hren, cpan, rzw, 230268160, wangdm, j.z.wang@seu.edu.cn).}%
\thanks{Yuqing Chen is with the Shenzhen International Graduate School, Tsinghua University, Beijing, China (email: chenyuqi24@mails.tsinghua.edu.cn).}%
}

\maketitle

\begin{abstract}
Integrated sensing and communication (ISAC) has emerged as a promising paradigm for future wireless networks. However, the extensive occupation of time-frequency resources for sensing  limits the communication capacity. To address the resource conflict, this paper proposes a novel \textit{sparse} sensing frame structure \textcolor{revisiondarkgreen}{aligned with the 5G new radio (5G NR) resource grid at the slot and physical resource block (PRB) levels through a predefined sensing-pattern block}. Specifically, inspired by the design of two-dimensional (2D) sparse antenna arrays, we map the  sparse spatial geometries onto the time-frequency resource grid. By leveraging the hole-free property of the difference coarray, we eliminate the range-velocity ambiguity and cast the sensing signals into a reconstructed \textcolor{revisiondarkgreen}{higher-order tensor} representation. For  efficient parameter estimation, we propose a novel \textcolor{revisiondarkgreen}{binary-tree-based tensor train decomposition (BT-TTD)} algorithm that extracts lower-dimensional \textcolor{revisiondarkgreen}{tensor train cores (TT-cores)} and executes a Vandermonde-prior-guided rectified
alternating least \textcolor{revisiondarkgreen}{squares} (RecALS) \textcolor{revisiondarkgreen}{algorithm}. Simulation results validate that the proposed scheme \textcolor{revisiondarkgreen}{achieves competitive estimation accuracy with reduced complexity}. Furthermore, we investigate the topological properties of various sparse patterns, revealing the inherent trade-off among sensing accuracy, resource occupancy, and computational complexity.
\end{abstract}


\section{Introduction}

\IEEEPARstart{T}{he} rapid proliferation of emerging applications, such as \textcolor{revisiondarkgreen}{immersive extended reality (XR)}, autonomous driving, and industrial internet of things (IIoT), imposes stringent requirements on future wireless infrastructure\cite{liu2022integrated,cui2021integrating}. These scenarios demand not only ubiquitous connectivity but also precise environmental awareness. In response to these challenges, integrated sensing and communication (ISAC) has emerged as a promising paradigm. By sharing the same spectrum and reusing the existing communication hardware, ISAC endows wireless systems with native sensing capabilities, thereby rendering it an indispensable element in the  future sixth-generation (6G) networks\cite{liu2020joint,lu2024integrated}.

As the dominant waveform in  current wireless networks (e.g., Wi-Fi and 5G), OFDM is envisioned as a prime candidate for future 6G networks\cite{zhu2009chunk}. This preference is attributed to its high spectral efficiency and robustness against multipath fading, as well as its orthogonal time-frequency structure that is intrinsically suitable for high-precision sensing tasks. Moreover, an OFDM-based ISAC framework is  compatible with existing communication standards, obviating the need for additional dedicated hardware architectures. As a result, extensive research efforts have been \textcolor{revisiondarkgreen}{devoted to OFDM-based ISAC, particularly in communication-centric scenarios}. 

For OFDM-based communication-centric ISAC, the randomness inherent in downlink data impairs the accuracy of target detection. This limitation is especially critical in \textcolor{revisiondarkgreen}{bistatic} scenarios, where the receiver operates without prior knowledge of the payload. Therefore, the design of dedicated sensing pilots  is essential for realizing wireless sensing. Notably, since the time-frequency structure of the dedicated pilots is known \textit{a priori} at the receiver, target information can be effectively retrieved by exploiting these pilots. To fully exploit the sensing potential, extensive studies have investigated the optimal placement of sensing pilots within the OFDM time-frequency resource grid \cite{mura2025optimized,li2025sensing,bicua2016generalized,zhang2024target,tang2021self,liu2025low,zhang2024ofdm,li2024mimo}. Generally, the exploration can be classified into two primary categories: waveform design and ambiguity mitigation. Specifically, the former aims to optimize transmitted waveforms for the sensing-related metrics, while the latter addresses the ambiguities arising from sparse frame structures, with the goal of extending the maximum unambiguous range for velocity and distance estimation.

In the field of waveform design, for instance, \cite{mura2025optimized} focused on minimizing the Cramér-Rao Bound (CRB) and employed  interpolation at the receiver to reconstruct the sparse sensing channel. Moreover, the authors of \cite{li2025sensing} considered multiple sensing metrics, including CRB, sidelobe levels, and sensing signal-to-noise ratio (SNR), to optimize the geometric pattern of transmit sensing resource elements (REs). However, these waveform-centric designs predominantly focused on the theoretical sensing potential of the transmit waveform, neglecting the practical processing chain required for OFDM receivers\cite{bicua2016generalized,zhang2024target}. In contrast, regarding the ambiguity mitigation, \cite{tang2021self} leveraged the varying frame lengths to extend the velocity detection range. Additionally, the authors of \cite{liu2025low} exploited the coprime property of the pilot intervals to eliminate the ambiguity. Furthermore, the authors of \cite{zhang2024ofdm} proposed a set of criteria for \textcolor{revisiondarkgreen}{the} placement of sensing REs, with the goal of maximizing the joint unambiguous range of distance and velocity estimation. Notably, while these methods can  achieve ambiguity performance equivalent to that of a fully allocated time-frequency grid theoretically, they typically rely on the traditional estimators (e.g., 2D-FFT\cite{zhang2024target} and 2D-MUSIC\cite{xie2021performance}), thereby suffering  from the limited resolution or prohibitive computational complexity. 

\textcolor{revisiondarkgreen}{Despite the aforementioned advances, it is noted that most existing works in both categories overlook the practical scheduling constraints of 5G new radio (5G NR) systems, particularly the use of slots and physical resource blocks (PRBs) as the basic scheduling units in the time and frequency domains, respectively.} Specifically, these approaches typically assume RE-level granularity for resource allocation, necessitating precise control over all individual REs. Such fine-grained flexibility, however, diverges from the standard specifications of 5G NR and inevitably incurs excessive control signaling overhead. Furthermore, the combined issues of high algorithmic complexity and  misalignment with practical OFDM receiver processing chains underscore the need for a more practical solution.

Motivated by  the above discussions, this paper develops a sensing frame structure compatible with the current 5G NR standards and proposes a tailored ultra-low complexity parameter extraction algorithm. The main contributions of this work  are summarized as follows:

\begin{itemize}
  \item We propose a novel sensing frame structure that is fully compatible with the 5G NR standard. Unlike existing methods that require fine-grained RE-level resource scheduling, our proposed scheme adopts \textcolor{revisiondarkgreen}{PRB-level granularity} and allows  flexible adjustment of the sensing resource overhead.
  
  \item Inspired by the design of 2D sparse antenna arrays, we map the spatial geometry of sparse arrays to the time-frequency resource grid for the sensing frame structure design. By leveraging the hole-free property of the difference \textcolor{revisiondarkgreen}{coarray} in 2D sparse arrays, we eliminate the range-velocity ambiguity caused by sparse RE placement and reformulate the sensing task as a parameter extraction problem based on the \textcolor{revisiondarkgreen}{autocorrelation} matrix.
  
  \item Subsequently,  we reconstruct the \textcolor{revisiondarkgreen}{autocorrelation matrix} into a higher-order tensor and embed shifting information into the higher dimensions to enhance parameter extraction performance. Furthermore, a \textcolor{revisiondarkgreen}{binary-tree-based tensor train decomposition (BT-TTD)} algorithm is developed to facilitate the rapid decomposition of the tensor.

  \item Finally, extensive simulations  demonstrate that the BT-TTD algorithm achieves  near-optimal parameter estimation accuracy with  substantially reduced complexity. Furthermore, by evaluating various sparse array topologies, we reveal an inherent three-way trade-off among sensing accuracy, resource occupancy, and computational complexity, thereby providing critical design guidelines for practical ISAC systems.

\end{itemize}

\section{Notation and Preliminaries}

In this section, \textcolor{revisiondarkgreen}{we provide preliminaries on notation and tensor operations}, including the \textcolor{revisiondarkgreen}{tensor train decomposition}, \textcolor{revisiondarkgreen}{canonical polyadic decomposition (CPD)}, and the properties of 2D sparse arrays.

\subsection{Notations}
Scalars, vectors, matrices, and tensors are respectively denoted by lowercase letters, boldface lowercase letters,  boldface uppercase letters, and calligraphic uppercase letters, e.g., $a$, $\mathbf{a}$, $\mathbf{A}$, and $\boldsymbol{\mathcal{A}}$. $\times_i$ denotes the mode-$i$ tensor product.  $(\cdot)^{\mathrm{T}}$, $(\cdot)^{\mathrm{H}}$, $(\cdot)^*$, and $(\cdot)^{\dagger}$ denote the transpose, the Hermitian transpose, the conjugation, and the pseudoinverse operators, respectively. $\mathbb{E}\{\cdot\}$ represents the statistical expectation. $\llbracket \cdot \rrbracket$ denotes the CPD representation of a tensor, while $\kappa(\cdot)$ denotes the \textcolor{revisiondarkgreen}{Kruskal rank}. The outer product, Kronecker product, Khatri-Rao product, and Hadamard product are denoted by $\circ$, $\otimes$, $\odot$, and $\diamond$, respectively. $\mathbf{X}_{(i)}$ denotes the mode-$i$ unfolding of a tensor $\boldsymbol{\mathcal{X}}$, and $[\cdot]_{\sqcup_i}$ denotes the tensor concatenation along the $i$-th dimension.  $\angle$ denotes the phase of a complex number. $|\cdot|$ denotes the cardinality of a set. 

\subsection{Tensor Preliminaries}
To facilitate understanding, we outline the basic definitions of \textcolor{revisiondarkgreen}{tensor networks}. \textcolor{revisiondarkgreen}{Readers are referred to} \cite{kolda2009tensor,oseledets2011tensor} for further details.

\begin{enumerate}[label=\textit{\arabic*})]
   \item \textit{Tensor train (TT) decomposition:} The TT decomposition represents a tensor $\boldsymbol{\mathcal{X}}\in \mathbb C^{I_1\times I_2\times \dots \times I_N}$ via a train of core tensors\cite{oseledets2011tensor}. Specifically, the $(i_1, \dots, i_D)$-th element  is given by   
    \begin{align}
         &~\quad\boldsymbol{\mathcal{X}}(i_1, \dots, i_D)= \boldsymbol{\mathcal{G}}_1(:, i_1, :) \boldsymbol{\mathcal{G}}_2(:, i_2, :) \cdots \boldsymbol{\mathcal{G}}_D(:, i_D, :) \nonumber\\
        &= \sum_{r_1, \dots, r_{D-1}=1}^{R_1, \dots, R_{D-1}} \mathbf{G}_1(i_1, r_1) \boldsymbol{\mathcal{G}}_2(r_1, i_2, r_2) \cdots \mathbf{G}_D(r_{D-1}, i_D),
    \end{align}where $\boldsymbol{{\mathcal{G}}}_d\in \mathbb C^{R_{d-1} \times I_d \times R_d}$ is the $d$-th TT-core with $R_0 = R_D = 1$. Finally, by collecting all entries,  we have
    \begin{equation}
        \boldsymbol{\mathcal{X}} = \mathbf{G}_1 \times_2^1 \boldsymbol{\mathcal{G}}_2 \times_3^1 \boldsymbol{\mathcal{G}}_3 \times_4^1 \cdots \times_{D}^{1} \mathbf{G}_D.
    \end{equation}
    
    \item \textit{Contraction product:} The $\times_n^m$ contraction product of $\boldsymbol{\mathcal{X}} \in \mathbb{C}^{I_1 \times \cdots \times I_N}$ and $\boldsymbol{\mathcal{Y}} \in \mathbb{C}^{J_1 \times \cdots \times J_M}$ (where $I_n = J_m$) produces an $(N+M-2)$-order tensor $\boldsymbol{\mathcal{Z}}$ \cite{biamonte2017tensor}. Specifically, we have
    \begin{equation}
        \begin{aligned}
        \boldsymbol{\mathcal{Z}} &= \boldsymbol{\mathcal{X}} \times_n^m \boldsymbol{\mathcal{Y}} \\
        &\in   \mathbb{R}^{I_1 \times \cdots \times I_{n-1} \times I_{n+1} \times \cdots \times I_N \times J_1 \times \cdots \times J_{m-1} \times J_{m+1} \times \cdots \times J_M},
        \end{aligned}
    \end{equation} with
    \begin{equation}
    \begin{aligned}
        \boldsymbol{\mathcal{Z}}&(i_1, \dots, i_{n-1}, i_{n+1}, \dots, i_N, j_1, \dots, j_{m-1}, j_{m+1}, \dots, j_M) \\
        &= \sum_{k=1}^{I_n} \boldsymbol{\mathcal{X}}(i_1, \dots, i_{n-1}, k, i_{n+1}, \dots, i_N) \\
        &\quad \times \boldsymbol{\mathcal{Y}}(j_1, \dots, j_{m-1}, k, j_{m+1}, \dots, j_M).
    \end{aligned}
    \end{equation}
    
    \item \textit{\textcolor{revisiondarkgreen}{CPD} and matrix unfolding:} The \textcolor{revisiondarkgreen}{CPD} factorizes $\boldsymbol{\mathcal{X}}\in \mathbb C^{I_1\times I_2\times \dots \times I_N}$ into a sum of rank-one tensors\cite{kolda2009tensor}. Specifically, with factor matrices $\{\mathbf{A}^{(n)}\}_{n=1}^N$ and a diagonal core $\boldsymbol{\boldsymbol{\mathcal{I}}}_{N,R}$, $\boldsymbol{\mathcal{X}}$ can be expressed as
    \begin{equation}
    \begin{aligned}
        \boldsymbol{\mathcal{X}} &= \sum_{r=1}^{R} \lambda_r \mathbf{a}_r^{(1)} \circ \mathbf{a}_r^{(2)} \circ \cdots \circ \mathbf{a}_r^{(N)} \\
        &= \boldsymbol{\boldsymbol{\mathcal{I}}}_{N,R} \times_1 \mathbf{A}^{(1)} \times_2 \mathbf{A}^{(2)} \cdots \times_N \mathbf{A}^{(N)},\\
        & \triangleq \llbracket \boldsymbol{\Lambda}; \mathbf{A}^{(1)}, \mathbf{A}^{(2)}, \dots, \mathbf{A}^{(N)} \rrbracket
    \end{aligned}
    \end{equation}
   where $\times_n$ denotes the mode-$n$ product and $\mathbf{A}^{(n)}=\left[ \mathbf{a}_1^{(n)},\mathbf{a}_2^{(n)},\dots,\mathbf{a}_R^{(n)}\right] \in \mathbb{C}^{I_n \times R},~ n=1\dots,N$ are the factor matrices. Consequently, the mode-$n$ unfolding of $\boldsymbol{\mathcal{X}}$ takes the following matrix form, e.g.,
    \begin{equation}
    \begin{aligned}
        \mathbf{X}_{(n)} = \mathbf{A}^{(n)} \mathbf{\Lambda} \Big( &\mathbf{A}^{(N)} \odot \cdots \odot \mathbf{A}^{(n+1)} \\
        &\odot \mathbf{A}^{(n-1)} \odot \cdots \odot \mathbf{A}^{(1)} \Big)^T,
    \end{aligned}
    \end{equation}
    where $\mathbf{\Lambda} = \text{diag}(\lambda_1, \dots, \lambda_R)$.
\end{enumerate}

\subsection{2D Sparse Arrays Preliminaries}

\begin{figure*}[htbp]
\centerline{\includegraphics[width=0.95\textwidth]{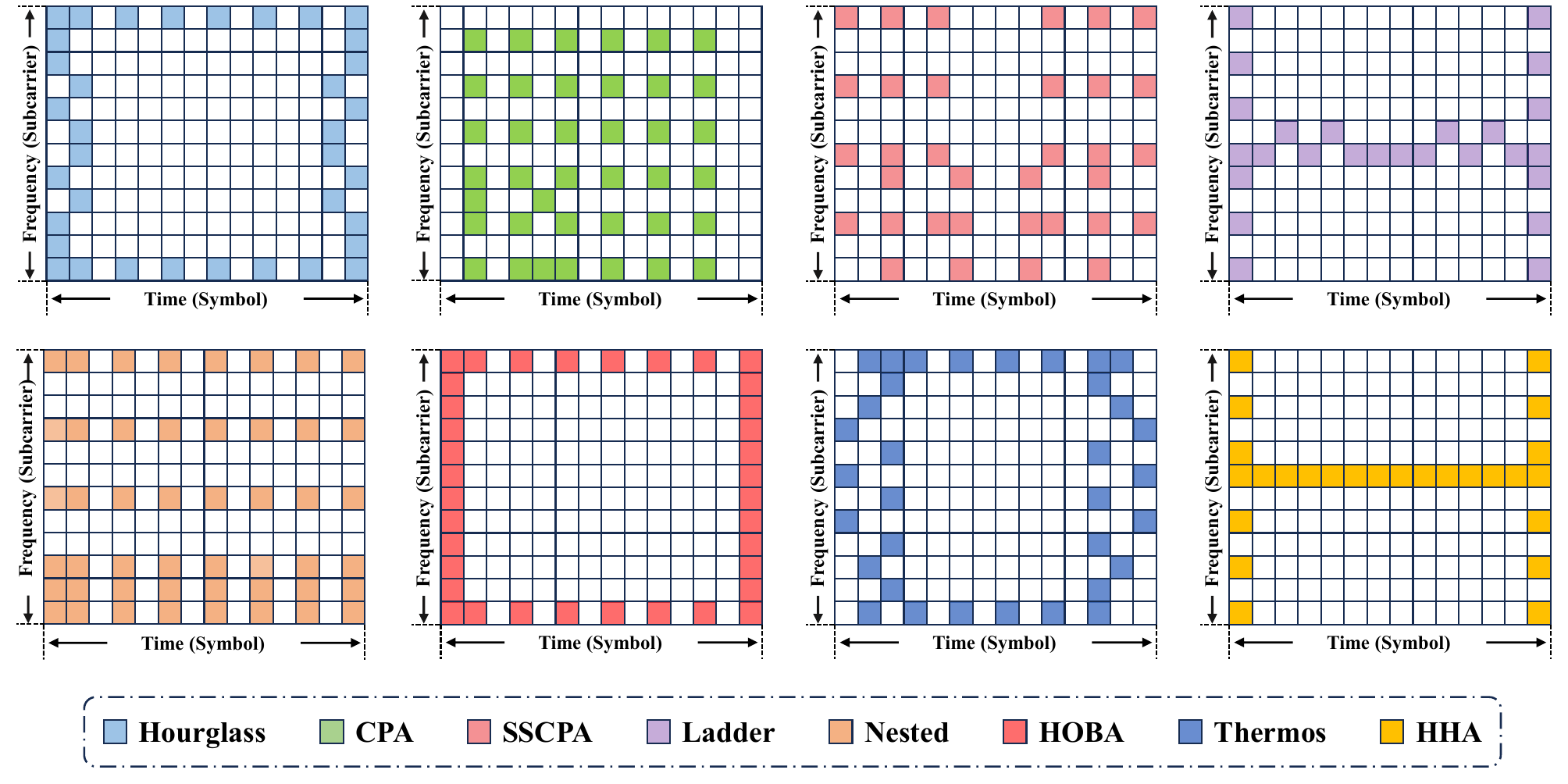}}
\captionsetup{labelfont={color=revisiondarkgreen},textfont={color=revisiondarkgreen}}
\caption{Examples of time-frequency resource allocation within one RB in an OFDM-ISAC system. The \textcolor{Color_Hourglass}{\rule{7pt}{7pt}}, \textcolor{Color_CPA}{\rule{7pt}{7pt}}, \textcolor{Color_SSCPA}{\rule{7pt}{7pt}}, \textcolor{Color_Ladder}{\rule{7pt}{7pt}}, \textcolor{Color_Nested}{\rule{7pt}{7pt}}, \textcolor{Color_HOBA}{\rule{7pt}{7pt}}, \textcolor{Color_Thermos}{\rule{7pt}{7pt}}, and \textcolor{Color_HHA}{\rule{7pt}{7pt}} blocks represent sensing resources arranged in Hourglass, CPA, SSCPA, Ladder, Nested, HOBA, Thermos, and HHA patterns, respectively, whereas the {\setlength{\fboxsep}{0pt}\setlength{\fboxrule}{0.4pt}\fcolorbox{black}{Color_Comm}{\textcolor{Color_Comm}{\rule{6.2pt}{6.2pt}}}} blocks denote communication resources.}
\label{fig:2D_Sparse_Array}
\end{figure*}

\begin{enumerate}[label=\textit{\arabic*})]

\item \textit{Difference coarray:} The difference coarray $\mathbb{D}$ of a 2D array $\mathbb{S}$ is defined as the set of \textcolor{revisiondarkgreen}{all pairwise differences among the elements of $\mathbb{S}$}, e.g.,
\begin{equation}
    \mathbb{D} = \left\{ \mathbf{n}_1 - \mathbf{n}_2 \mid \mathbf{n}_1, \mathbf{n}_2 \in \mathbb{S} \right\},
\end{equation}
where $\mathbf{n} \in \mathbb{Z}^2$ is the integer-valued position vector.

\item \textit{Uniform degrees of \textcolor{revisiondarkgreen}{freedom} (uDOFs) and hole-free property:} Let $\mathbb{U}$ denote the maximum central rectangular part of  $\mathbb{D}$. \textcolor{revisiondarkgreen}{The number of uDOFs is defined as the cardinality} $|\mathbb{U}|$. Furthermore, an array is  hole-free if 
    \begin{equation}
        \mathbb{D} = \mathbb{U}.
    \end{equation}

    \item \textcolor{revisiondarkgreen}{\textit{Examples of known 2D sparse arrays:} To illustrate practical configurations, Fig.~\ref{fig:2D_Sparse_Array} depicts the allocation of sensing REs within one RB\footnote{Throughout this paper, RB denotes the proposed sensing-pattern block rather than the conventional 4G resource block.} following eight representative geometries: hourglass array (HA) and half-open-box array (HOBA)\cite{liu2017hourglass}, coprime planar array (CPA)\cite{vaidyanathan2010sparse}, shift-based symmetric coprime planar array (SSCPA)\cite{zhang2025shift}, 2D nested array (Nested)\cite{pal2012nested}, half-H array (HHA) and ladder array (LA)\cite{ren2018extensions}, and thermos array (Thermos)\cite{sun2018thermos}. We refer readers to \cite{liu2017hourglass,vaidyanathan2010sparse,zhang2025shift,pal2012nested,ren2018extensions,sun2018thermos} for detailed design methodologies and properties.}

\end{enumerate}

\section{System Model}

We consider a bistatic OFDM-ISAC system, where the transmitter (Tx) leverages the inherent flexibility of OFDM resource allocation to map the sensing and communication signals onto distinct REs. Simultaneously, the receiver (Rx) captures the downlink communication stream concurrently with the echo signals reflected from the neighboring targets. To prioritize the analysis of frame structure design, we assume that  the Tx and Rx are \textcolor{revisiondarkgreen}{equipped with a single antenna}.\footnote{This assumption aligns with the standard practice in frame structure design literature \cite{li2025sensing, gong2025frame,liu2025low,zhang2024ofdm,mura2025optimized}. Note that the proposed frame structure framework can be readily \textcolor{revisiondarkgreen}{extended to MIMO architectures}.}

\subsection{Transmit Signal Model}
Let the transmit signal consist of $N$ subcarriers and $M$ OFDM symbols. To avoid the interference between  sensing and communication functions, we employ an orthogonal resource allocation strategy. Specifically, we define two binary RE selection indicator matrices $\mathbf{U}_r \in \{0,1\}^{N \times M}$ and $\mathbf{U}_c \in \{0,1\}^{N \times M}$, i.e.,
\begin{equation}
    [\mathbf{U}_r]_{n,m} = 
    \begin{cases} 
    1, & \text{if the } (n,m)\text{-th RE is allocated for sensing,} \\
    0, & \text{otherwise,}
    \end{cases}
\end{equation}
and
\begin{equation}
    [\mathbf{U}_c]_{n,m} = 
    \begin{cases} 
    1, & \text{if the } (n,m)\text{-th RE is for communication,} \\
    0, & \text{otherwise.}
    \end{cases}
\end{equation}
Here,  we set $\mathbf{U}_r \diamond \mathbf{U}_c = \mathbf{0}$. Assuming that the same predefined sensing pattern is repeated over $F$ frequency-block positions and $S$ time-block positions, we have
$$\mathbf{U}_r = \mathbf{J}_{F, S} \otimes \mathbf{U}_{\text{RB}}$$
\textcolor{revisiondarkgreen}{where $\mathbf J_{F,S}\in\mathbb R^{F\times S}$ denotes the all-ones matrix, and $\mathbf U_{\rm RB}\in\{0,1\}^{N_{\rm RB}\times M_{\rm RB}}$ specifies the sensing-RE pattern within a predefined RB aligned with the NR PRB and slot boundaries. Here, $F$ and $S$ denote the numbers of block repetitions in the frequency and time domains, respectively. The resulting frame comprises $N=FN_{\rm RB}$ subcarriers, $M=SM_{\rm RB}$ OFDM symbols, and $N_{\rm snap}=FS$ sensing snapshots.}

Then, let $\mathbf{S}_r \in \mathbb{C}^{N \times M}$ and $\mathbf{S}_c \in \mathbb{C}^{N \times M}$ denote the sensing and communication symbol matrices, respectively.  The frequency-domain OFDM transmit signal matrix $\mathbf{X} \in \mathbb{C}^{N \times M}$ can be formulated as
\begin{equation}
\begin{aligned}
    \mathbf{X} &= \mathbf{U}_r \diamond \mathbf{S}_r + \mathbf{U}_c \diamond \mathbf{S}_c \\
    &= \left( \mathbf{J}_{F, S} \otimes \mathbf{U}_{\text{RB}} \right) \diamond \mathbf{S}_r + \mathbf{U}_c \diamond \mathbf{S}_c.
\end{aligned}
\end{equation}

\noindent Correspondingly, the time-domain baseband OFDM signal can be expressed as
\begin{equation}
    x(t) = \sum_{m=0}^{M-1} \sum_{n=0}^{N-1} [\mathbf{X}]_{n, m} e^{j 2\pi n \Delta f (t - m T_{\text{sym}} - T_{\text{cp}})} g(t - m T_{\text{sym}}),
\end{equation}
where $\Delta f$ is the subcarrier spacing, $T_{\text{cp}}$ is the CP duration, and $T_{\text{sym}} = 1/\Delta f + T_{\text{cp}}$ is the total symbol duration. $g(t)$ denotes the rectangular pulse shaping window of width $T_{\text{sym}}$.

\subsection{Received signal for sensing}

 \textcolor{revisiondarkgreen}{For $Q$ point targets}, the frequency domain echo signal on the $(n,m)$-th RE can be expressed as\footnote{It is worth noting that the symbol timing offset (STO) and carrier frequency offset (CFO) between the Tx and the Rx introduce constant biases into the range and velocity estimation. To address this, dedicated synchronization algorithms can be employed to  calibrate and compensate the errors \cite{ni2021uplink, wu2024sensing}.  With a primary focus on the sensing frame structure design, we follow the convention in the existing literature \cite{kumari2019adaptive, zhang2024ofdm} by assuming perfect synchronization, i.e., neglecting the STO and CFO between them.}
\begin{equation}
     [\mathbf Y]_{n,m} = \sum_{q=1}^{Q} \beta_q   [\mathbf X]_{n,m} e^{-j 2\pi n \Delta f \tau_q} e^{j 2\pi m T_{\text{sym}} f_{D,q}} +  [\mathbf{Z}]_{n,m},
\end{equation}
where $\beta_q$ denotes the complex attenuation coefficient of the $q$-th target, $\tau_q$ and $f_{D,q}$ denote the associated time delay and Doppler shift, and  $ [\mathbf{Z}]_{n,m} \sim \mathcal{CN}(0, \sigma^2)$  denotes the additive white Gaussian noise (AWGN).

In the considered bistatic sensing scenario, the communication payload carried by non-sensing REs is unknown to the receiver. Therefore, we exclusively utilize the sensing REs for parameter estimation. Following the standard OFDM radar processing chain\cite{li2025sensing,zhang2024target}, we perform element-wise division of the received signal by the known sensing symbols $\mathbf{S}_r$. Consequently, the resulting \textcolor{revisiondarkgreen}{frequency-domain sensing channel} is expressed as
\begin{equation}
\begin{aligned}
        \mathbf{H}_r&= \mathbf{U}_r \diamond \left( \sum_{q=1}^{Q} \beta_q \mathbf{a}_{\tau}(\tau_q)  \mathbf{a}_{D}^H(f_{D,q}) \right) + \tilde{\mathbf{Z}},\\
         &= \left(\mathbf{J}_{F, S} \otimes  \mathbf{U}_\text{RB}\right) \diamond \left( \sum_{q=1}^{Q} \beta_q \mathbf{a}_{\tau}(\tau_q)  \mathbf{a}_{D}^H(f_{D,q}) \right)+\tilde{\mathbf{Z}}\textcolor{revisiondarkgreen}{,}
\end{aligned}
\end{equation}
{\color{revisiondarkgreen}
where $\tilde{\mathbf Z}\in\mathbb C^{N\times M}$ denotes the processed noise term.
The delay and Doppler steering vectors over the complete OFDM grid are defined as
\begin{equation}
\begin{aligned}
\boldsymbol{a}_{\tau}\left(\tau_{q}\right) &=\left[1, e^{-\jmath 2 \pi \Delta{f} \tau_{q}}, \ldots, e^{-\jmath 2 \pi(N-1) \Delta{f} \tau_{q}}\right]^{T}, \\
\boldsymbol{a}_{D}\left(f_{D, q}\right)&=\left[1, e^{-\jmath 2 \pi f_{D, q} T_{\mathrm{sym}}}, \ldots, e^{-\jmath 2 \pi(M-1) f_{D, q} T_{\mathrm{sym}}}\right]^{T}.
\end{aligned}
\end{equation}
To explicitly characterize the observation associated with each repeated sensing-pattern block, let $(f,s)$ denote its frequency--time block position, where $f=0,\ldots,F-1$ and $s=0,\ldots,S-1$. For a local RE indexed by $(u,v)$ within the block, the corresponding global RE indices are
\begin{equation}\label{eq:global_local_indices}
    n=fN_{\rm RB}+u,\qquad m=sM_{\rm RB}+v,
\end{equation}
where $u=0,\ldots,N_{\rm RB}-1$ and $v=0,\ldots,M_{\rm RB}-1$. Accordingly, the effective coefficient of the $q$-th target in the $(f,s)$-th block is
\begin{equation}\label{eq:beta_block}
    \beta_{q,f,s}
    =
    \beta_q
    e^{-j2\pi fN_{\rm RB}\Delta f\tau_q}
    e^{j2\pi sM_{\rm RB}T_{\rm sym}f_{D,q}}.
\end{equation}
The corresponding block-level sensing channel can then be expressed as
\begin{equation}\label{eq:H_RB}
\begin{aligned}
    \mathbf H_{{\rm RB},f,s}
    &=
    \mathbf U_{\rm RB}\diamond
    \left(
    \sum_{q=1}^{Q}
    \beta_{q,f,s}
    \mathbf a_{{\rm RB},\tau}(\tau_q)
    \mathbf a_{{\rm RB},D}^{H}(f_{D,q})
    \right)\\
    &\quad +\widetilde{\mathbf Z}_{{\rm RB},f,s},
\end{aligned}
\end{equation}
where $\widetilde{\mathbf Z}_{{\rm RB},f,s}\in\mathbb C^{N_{\rm RB}\times M_{\rm RB}}$ denotes the processed noise within the $(f,s)$-th block. The corresponding within-block delay and Doppler steering vectors are defined as
\begin{equation}
\begin{aligned}
\mathbf a_{{\rm RB},\tau}(\tau_q)
&=\left[1,e^{-\jmath2\pi\Delta f\tau_q},\ldots,e^{-\jmath2\pi(N_{\rm RB}-1)\Delta f\tau_q}\right]^T,\\
\mathbf a_{{\rm RB},D}(f_{D,q})
&=\left[1,e^{-\jmath2\pi T_{\rm sym}f_{D,q}},\ldots,e^{-\jmath2\pi(M_{\rm RB}-1)T_{\rm sym}f_{D,q}}\right]^T.
\end{aligned}
\end{equation}
}

\subsection{CRB Analysis on the Resource Allocation}

Before detailing the proposed frame structure, we analyze the OFDM resource allocation via the CRB to reveal how the time-frequency resource distribution affects estimation accuracy. Without loss of generality, we consider a single-target scenario, i.e., $Q=1$, and introduce the following theorem to characterize the  estimation limits.

{\color{revisiondarkgreen}
\itshape \textbf{Theorem 1:} Consider a single-target scenario and let $\mathbb{S}_{\rm frame}\triangleq\{(n,m)\mid[\mathbf U_r]_{n,m}=1\}=\{(n_k,m_k)\}_{k=1}^{K_{\rm frame}}$ denote the frame-level sensing-RE set, where $K_{\rm frame}=|\mathbb{S}_{\rm frame}|$ is the total number of sensing REs in the OFDM frame. Let $V_n$, $V_m$, and $C_{nm}$ denote the population variance of $\{n_k\}$, the population variance of $\{m_k\}$, and their covariance, respectively. Define the normalized coupling coefficient as
\begin{equation}
    \rho=\frac{C_{nm}}{\sqrt{V_nV_m}},
    \qquad |\rho|\leq1.
\end{equation}
For $V_n>0$, $V_m>0$, and $|\rho|<1$, the exact CRB matrix for $\boldsymbol{\xi}=[\tau,f_D]^T$ is
\begin{equation}
\operatorname{CRB}(\boldsymbol{\xi})
=
\kappa_{\rho}
\begin{bmatrix}
\dfrac{1}{\Delta f^2V_n}
&
\dfrac{\rho}{\Delta fT_{\mathrm{sym}}\sqrt{V_nV_m}}
\\[2mm]
\dfrac{\rho}{\Delta fT_{\mathrm{sym}}\sqrt{V_nV_m}}
&
\dfrac{1}{T_{\mathrm{sym}}^2V_m}
\end{bmatrix},
\end{equation}
where
\begin{equation}
    \kappa_{\rho}
    =
    \frac{\sigma_z^2}{8\pi^2|\beta|^2K_{\rm frame}(1-\rho^2)},
\end{equation}
and $\sigma_z^2$ is the noise variance. The corresponding marginal CRBs are
\begin{subequations}
\begin{align}
    \operatorname{CRB}(\tau)
    &=
    \frac{\sigma_z^2}{8\pi^2|\beta|^2K_{\rm frame}\Delta f^2V_n(1-\rho^2)},
    \label{eq:crb_range}\\
    \operatorname{CRB}(f_D)
    &=
    \frac{\sigma_z^2}{8\pi^2|\beta|^2K_{\rm frame}T_{\mathrm{sym}}^2V_m(1-\rho^2)}.
    \label{eq:crb_velocity}
\end{align}
\end{subequations}
}
\par\noindent\itshape \textbf{Proof:}  \upshape Please refer to Appendix~\ref{app:proof_theorem_1}. \hfill $\blacksquare$
\par\vspace{\baselineskip}

\textcolor{revisiondarkgreen}{From Theorem 1, we obtain the insight that minimizing the delay and Doppler CRBs requires maximizing $V_n(1-\rho^2)$ and $V_m(1-\rho^2)$, respectively. Therefore, an effective resource-allocation strategy should place the sensing REs near the edges of the resource grid to increase their spread in both the frequency and time domains. Moreover, it should maintain a low correlation between the frequency and time indices to mitigate delay--Doppler coupling.}\footnote{Note that in future communication systems, communication pilots (e.g., DMRS or CSI-RS) could potentially be reused as sensing resources. \textcolor{revisiondarkgreen}{The joint optimization of these resources, however, falls outside} the scope of this paper, which is dedicated to the sensing frame structure design.}

\section{Tensor Derivation for Sparse Sensing REs}
\label{sec:tensor_derivation}

In this section, we first convert the sparse sensing REs into \textcolor{revisiondarkgreen}{an equivalent virtual uniform rectangular array (URA)}. Subsequently, the \textcolor{revisiondarkgreen}{tensor dimension-increment technique} is employed to reconstruct the corresponding 2D \textcolor{revisiondarkgreen}{autocorrelation matrix} into a 4D tensor, thereby enabling us to fully leverage the delay-Doppler shifting information embedded within the \textcolor{revisiondarkgreen}{virtual URA} for high-resolution estimation.

\subsection{\textcolor{revisiondarkgreen}{Autocorrelation Analysis and Virtual URA Construction}}
\label{sec:virtual URA Construction}
{\color{revisiondarkgreen}
Following \cite{xie2021performance}, we exploit the repeated sensing-pattern blocks to obtain the observations required for 2D smoothing. Let
\begin{equation}
\begin{aligned}
    \mathbb S_{\rm SPB}
    &\triangleq\left\{(u,v)\mid[\mathbf U_{\rm RB}]_{u,v}=1\right\}\\
    &=\left\{(u_k,v_k)\right\}_{k=1}^{K_{\rm SPB}}
\end{aligned}
\end{equation}
denote the set of active sensing-RE positions within one sensing-pattern block, with $K_{\rm SPB}=|\mathbb S_{\rm SPB}|$. Under the global-to-local mapping in \eqref{eq:global_local_indices}, $\mathbb S_{\rm frame}$ is the union of the $FS$ shifted blocks corresponding to $\mathbb S_{\rm SPB}$, with $K_{\rm frame}=FSK_{\rm SPB}$. For the $(f,s)$-th block, the corresponding active sensing observations are extracted from $\mathbf H_{{\rm RB},f,s}$ in \eqref{eq:H_RB} and vectorized as
\begin{equation} \label{eq:h_vec}
\begin{aligned}
    \mathbf h_{{\rm RB},f,s}
    &=\mathbf E_{\mathbb S_{\rm SPB}}\operatorname{vec}\!\left(\mathbf H_{{\rm RB},f,s}\right)\\
    &=\mathbf A_{\mathbb S_{\rm SPB}}\boldsymbol\beta_{f,s}
    +\mathbf z_{{\rm RB},f,s}
    \in\mathbb C^{K_{\rm SPB}\times1},
\end{aligned}
\end{equation}
where $\mathbf E_{\mathbb S_{\rm SPB}}\in\{0,1\}^{K_{\rm SPB}\times N_{\rm RB}M_{\rm RB}}$ extracts the active sensing REs, $\boldsymbol\beta_{f,s}=[\beta_{1,f,s},\ldots,\beta_{Q,f,s}]^T$, and $\mathbf z_{{\rm RB},f,s}=\mathbf E_{\mathbb S_{\rm SPB}}\operatorname{vec}(\widetilde{\mathbf Z}_{{\rm RB},f,s})$. The effective manifold matrix $\mathbf A_{\mathbb S_{\rm SPB}}\in\mathbb C^{K_{\rm SPB}\times Q}$ is
\begin{equation}
    \mathbf A_{\mathbb S_{\rm SPB}}
    =\left[
    \mathbf a_{\mathbb S_{\rm SPB}}(\tau_1,f_{D,1}),\ldots,
    \mathbf a_{\mathbb S_{\rm SPB}}(\tau_Q,f_{D,Q})
    \right],
\end{equation}
where $\mathbf a_{\mathbb S_{\rm SPB}}(\tau_q,f_{D,q})=\allowbreak\mathbf E_{\mathbb S_{\rm SPB}}\bigl[\mathbf a_{{\rm RB},D}^{*}(f_{D,q})\otimes\allowbreak\mathbf a_{{\rm RB},\tau}(\tau_q)\bigr]$ is the effective steering vector of the $q$-th target.
Each vector $\mathbf h_{{\rm RB},f,s}$ constitutes one sensing snapshot, yielding $N_{\rm snap}=FS$ snapshots within the considered OFDM frame.
}

{\color{revisiondarkgreen}
    Assuming that $\mathbf z_{{\rm RB},f,s}$ is zero mean, independent of the target echoes, and satisfies $\mathbb E_{\mathbf z}[\mathbf z_{{\rm RB},f,s}\mathbf z_{{\rm RB},f,s}^{H}]=\sigma_z^2\mathbf I_{K_{\rm SPB}}$, the population autocorrelation matrix is
\begin{equation}\label{eq:R_RB_population}
\begin{aligned}
    \mathbf R_{\rm RB}
    &\triangleq
    \mathbb E_{f,s}\!\left[
    \mathbb E_{\mathbf z}\!\left[
    \mathbf h_{{\rm RB},f,s}\mathbf h_{{\rm RB},f,s}^{H}
    \right]\right]\\
    &=\mathbf A_{\mathbb S_{\rm SPB}}\mathbf R_{\beta,{\rm RB}}\mathbf A_{\mathbb S_{\rm SPB}}^{H}
    +\sigma_z^2\mathbf I_{K_{\rm SPB}},
\end{aligned}
\end{equation}
where $\mathbb E_{f,s}[\,\cdot\,]\triangleq (FS)^{-1}\sum_{f=0}^{F-1}\sum_{s=0}^{S-1}(\,\cdot\,)$, and the block-position-averaged target-coefficient correlation matrix is
\begin{equation}\label{eq:R_beta_RB}
    \mathbf R_{\beta,{\rm RB}}
    \triangleq
    \mathbb E_{f,s}
    \left[\boldsymbol\beta_{f,s}\boldsymbol\beta_{f,s}^{H}\right].
\end{equation}
\par\indent To justify the diagonal target-power representation used in the subsequent virtual coarray construction, we examine the off-diagonal entries of $\mathbf R_{\beta,{\rm RB}}$. The following proposition provides a sufficient condition under which averaging over the repeated block positions suppresses the inter-target cross terms.

{\itshape\textbf{Proposition 1:} Define $\Delta\tau_{q,\ell}=\tau_q-\tau_\ell$ and $\Delta f_{D,q,\ell}=f_{D,q}-f_{D,\ell}$. If the inter-block phase sequences satisfy
\begin{equation}\label{eq:dirichlet_condition}
\begin{aligned}
\epsilon
&\triangleq
\max_{q\neq\ell}
\left|
D_F\!\left(2\pi N_{\rm RB}\Delta f\Delta\tau_{q,\ell}\right)
\right.\\
&\qquad\left.
{}\times D_S\!\left(2\pi M_{\rm RB}T_{\rm sym}\Delta f_{D,q,\ell}\right)
\right|
\ll1,
\end{aligned}
\end{equation}
where $D_L(x)\triangleq L^{-1}\sum_{k=0}^{L-1}e^{\jmath kx}$ denotes the normalized Dirichlet kernel, then
\begin{equation}\label{eq:R_beta_approx}
    \mathbf R_{\beta,{\rm RB}}
    =\mathbf R_s+\boldsymbol\Delta_\beta,
    \qquad
    \mathbf R_s
    =\operatorname{diag}\!\left(|\beta_1|^2,\ldots,|\beta_Q|^2\right),
\end{equation}
where $[\boldsymbol\Delta_\beta]_{q,q}=0$ and the off-diagonal entries satisfy
\begin{equation}\label{eq:R_beta_error_bound}
    \left|[\boldsymbol\Delta_\beta]_{q,\ell}\right|
    \leq\epsilon|\beta_q||\beta_\ell|,
    \qquad q\neq\ell.
\end{equation}
Consequently, $\mathbf R_{\beta,{\rm RB}}\approx\mathbf R_s$ when $\epsilon\ll1$, with exact equality when $\epsilon=0$.}
\par\noindent\textit{\textbf{Proof:}} Please refer to Appendix~\ref{app:proof_proposition_1}. \hfill $\blacksquare$
\par\vspace{\baselineskip}

Under the condition in Proposition 1, \eqref{eq:R_RB_population} becomes
\begin{equation}\label{eq:R_RB_approx}
    \mathbf R_{\rm RB}
    \approx
    \mathbf A_{\mathbb S_{\rm SPB}}\mathbf R_s\mathbf A_{\mathbb S_{\rm SPB}}^{H}
    +\sigma_z^2\mathbf I_{K_{\rm SPB}}.
\end{equation}
Then, by vectorizing \eqref{eq:R_RB_approx}, we have
\begin{equation}
\begin{aligned}
    \boldsymbol r_{\rm RB}
    &=\operatorname{vec}\!\left(\mathbf R_{\rm RB}\right)\\
    &\approx\sum_{q=1}^{Q}|\beta_q|^2
    \Bigl(
    \mathbf a_{\mathbb S_{\rm SPB}}^{*}(\tau_q,f_{D,q})
    \otimes\mathbf a_{\mathbb S_{\rm SPB}}(\tau_q,f_{D,q})
    \Bigr)\\
    &\quad+\sigma_z^2\operatorname{vec}(\mathbf I_{K_{\rm SPB}}),
\end{aligned}\label{eq:vec_R}
\end{equation}
where $|\beta_q|^2$ denotes the power of the $q$-th target.
}
{\color{revisiondarkgreen}
The Kronecker product $\mathbf{a}_{\mathbb{S}_{\rm SPB}}^*(\tau_q,f_{D,q})\otimes\mathbf{a}_{\mathbb{S}_{\rm SPB}}(\tau_q,f_{D,q})$ in \eqref{eq:vec_R} acts as a virtual steering vector indexed by the block-level difference coarray
\begin{equation}
    \mathbb{D}_{\rm SPB}
    =\left\{(u_i-u_j,v_i-v_j)\mid
    (u_i,v_i),(u_j,v_j)\in\mathbb{S}_{\rm SPB}\right\}.
\end{equation}
Each $(p,q)\in\mathbb D_{\rm SPB}$ specifies a frequency-domain lag $p$ and a symbol-domain lag $q$. We then extract the largest origin-centered contiguous rectangular subset of $\mathbb D_{\rm SPB}$ as
\begin{equation}
\begin{aligned}
    \mathbb U&=\mathcal P\times\mathcal Q\subseteq\mathbb D_{\rm SPB},\\
    \mathcal P&=\{p_{\min},p_{\min}+1,\ldots,p_{\max}\},\\
    \mathcal Q&=\{q_{\min},q_{\min}+1,\ldots,q_{\max}\}.
\end{aligned}
\end{equation}
This subset forms a virtual URA of size $M_v\times N_v$, where $M_v=|\mathcal P|$ and $N_v=|\mathcal Q|$. For each lag $(p,q)\in\mathbb U$, the redundant autocorrelation entries are averaged as
\begin{equation}
    r_{\mathbb D_{\rm SPB}}(p,q)
    =\frac{1}{w(p,q)}
    \sum_{\substack{i,j:\\
    (u_i-u_j,v_i-v_j)=(p,q)}}
    [\mathbf R_{\rm RB}]_{i,j},
\end{equation}
where $w(p,q)$ denotes the number of sensing-RE pairs producing the lag $(p,q)$. The virtual signal matrix is then formed as
\begin{equation}
    [\mathbf U_{\rm vir}]_{a,b}
    =r_{\mathbb D_{\rm SPB}}\!\left(p_{\min}+a-1,q_{\min}+b-1\right),
\end{equation}
for $a=1,\ldots,M_v$ and $b=1,\ldots,N_v$. Accordingly, $\mathbf U_{\rm vir}$ can be expressed in terms of the virtual steering vectors as
\begin{equation}
    \mathbf U_{\rm vir}
    =\sum_{q=1}^{Q}|\beta_q|^2
    \mathbf b_{\tau}(\tau_q)\circ\mathbf b_D^*(f_{D,q})
    +\mathbf Z_{\rm vir}
    \in\mathbb C^{M_v\times N_v},
\end{equation}
where $\mathbf Z_{\rm vir}\in\mathbb C^{M_v\times N_v}$ represents the mapped noise term, and the virtual steering vectors are
\begin{equation}
\begin{aligned}
    \mathbf b_{\tau}(\tau_q)
    &=\left[e^{-j2\pi\Delta f p_{\min}\tau_q},\ldots,
    e^{-j2\pi\Delta f p_{\max}\tau_q}\right]^T,\\
    \mathbf b_D(f_{D,q})
    &=\left[e^{-j2\pi T_{\rm sym}q_{\min}f_{D,q}},\ldots,
    e^{-j2\pi T_{\rm sym}q_{\max}f_{D,q}}\right]^T.
\end{aligned}
\end{equation}
}

\textbf{\itshape Remark 1:} Directly applying subspace decomposition to
$\mathbf R_{\rm RB}$ is equivalent to the standard 2D-MUSIC algorithm
\cite{xie2021performance}. However, the 2D-MUSIC estimator fails to exploit the
structural characteristics of the virtual URA $\mathbb U$ derived from the
difference coarray $\mathbb D_{\rm SPB}$. Consequently, the inherent uDOFs
offered by the virtual array are not fully utilized, leading to a loss in
estimation performance.

\subsection{\textcolor{revisiondarkgreen}{Structured 4D Tensor Construction via Tensor Dimension Increment}}

In this subsection, we adopt the \textcolor{revisiondarkgreen}{tensor dimension-increment technique} to reformulate the virtual signal matrix $\mathbf{U}_{\text{vir}}$ into a \textcolor{revisiondarkgreen}{higher-order tensor} for parameter estimation\cite{zheng2023coarray}. Specifically,  by applying a segmentation window of size $P_\tau \times P_D$, the $(l_\tau, l_D)$-th sub-matrix of $\mathbf{U}_{\text{vir}}$ is given by
\begin{equation}
    \mathbf{U}_{(l_\tau, l_D)} = \mathbf{U}_{\text{vir}}(l_\tau : l_\tau + P_\tau - 1, \, l_D : l_D + P_D - 1) \in \mathbb{C}^{P_\tau \times P_D},
\end{equation}
where  $l_\tau = 1, \dots, L_\tau$ and $l_D = 1, \dots, L_D$, with $L_\tau = M_v - P_\tau + 1$ and $L_D = N_v - P_D + 1$. To ensure that the number of sub-matrices is sufficient for CPD, the window sizes are constrained by 
\begin{equation}
    2 \le P_\tau \le M_v - 1, \quad 2 \le P_D \le N_v - 1.
\end{equation}

\noindent It is readily verified that the sub-matrices $\mathbf{U}_{(l_\tau, l_D)}$ share the same delay and Doppler structure but exhibit a linear phase shift across the sliding indices.  By leveraging this shifting invariance property, we can construct a \textcolor{revisiondarkgreen}{higher-order tensor} where the shifting information is explicitly embedded into additional tensor dimensions.

First, for a fixed $l_D$, we stack the sub-matrices along the third dimension to construct a \textcolor{revisiondarkgreen}{3D} tensor embedded with the delay-domain shifting information
\begin{equation}
\begin{split}
    \boldsymbol{\mathcal{I}}_{l_D} &= \left[ \mathbf{U}_{(1, l_D)}, \mathbf{U}_{(2, l_D)}, \cdots, \mathbf{U}_{(L_\tau, l_D)} \right]_{\sqcup_3} \\
    &\in \mathbb{C}^{P_\tau \times P_D \times L_\tau}, \quad \forall l_D = 1,\dots, L_D.
\end{split}
\end{equation}

\noindent Subsequently, the \textcolor{revisiondarkgreen}{3D} tensors $\left\{\boldsymbol{\mathcal{I}}_{l_D}, l_{D}=1,2, \ldots, L_{D}\right\}$ are concatenated along the fourth dimension to construct the final \textcolor{revisiondarkgreen}{4D} tensor embedded with the Doppler-domain shifting information
\begin{equation}
\begin{split}
    \boldsymbol{\mathcal{T}} &= \left[ \boldsymbol{\mathcal{I}}_{1}, \dots, \boldsymbol{\mathcal{I}}_{L_D} \right]_{\sqcup_4} \in \mathbb{C}^{P_\tau \times P_D \times L_\tau \times L_D} \\
    &= \sum_{q=1}^{Q} |\beta_q|^2 \, \mathbf{b}_{\text{sub},\tau}(\tau_q) \circ \mathbf{b}_{\text{sub},D}(f_{D,q}) \\
    &\quad \circ \mathbf{b}_{\text{shift},\tau}(\tau_q) \circ \mathbf{b}_{\text{shift},D}(f_{D,q}) + \boldsymbol{\mathcal{Z}} \\
    &\triangleq \llbracket \mathbf{\Lambda}_{|\beta|^2}; \mathbf{B}_{\text{sub},\tau}, \mathbf{B}_{\text{sub},D}, \mathbf{B}_{\text{shift},\tau}, \mathbf{B}_{\text{shift},D} \rrbracket + \boldsymbol{\mathcal{Z}},
\end{split}\label{eq:4D_Tensor}
\end{equation}
where \textcolor{revisiondarkgreen}{$\mathbf{\Lambda}_{|\beta|^2}\triangleq\mathbf R_s$ denotes the diagonal target-power matrix defined in \eqref{eq:R_beta_approx}}, and
\begin{equation}
  \begin{aligned}
    \mathbf{b}_{\text{sub},\tau}(\tau_q) &= \left[ 1, e^{-j 2\pi \Delta f \tau_q}, \dots, e^{-j 2\pi \Delta f (P_\tau-1) \tau_q} \right]^T \in \mathbb{C}^{P_\tau}, \\
    \mathbf{b}_{\text{sub},D}(f_{D,q}) &= \left[ 1, e^{j 2\pi T_{\text{sym}} f_{D,q}}, \dots, e^{j 2\pi T_{\text{sym}} (P_D-1) f_{D,q}} \right]^T \in \mathbb{C}^{P_D}, 
\end{aligned}  
\end{equation}
denote the delay and Doppler factors in the sub-matrices, and
\begin{equation}
  \begin{aligned}
    \mathbf{b}_{\text{shift},\tau}(\tau_q) &= \left[ 1, e^{-j 2\pi \Delta f \tau_q}, \dots, e^{-j 2\pi \Delta f (L_\tau-1) \tau_q} \right]^T \in \mathbb{C}^{L_\tau}, \\
    \mathbf{b}_{\text{shift},D}(f_{D,q}) &= \left[ 1, e^{j 2\pi T_{\text{sym}} f_{D,q}}, \dots, e^{j 2\pi T_{\text{sym}} (L_D-1) f_{D,q}} \right]^T \in \mathbb{C}^{L_D},
\end{aligned}  
\end{equation}
denote the shifting factors corresponding to the shifting information embedded in the 4D tensor construction. Furthermore, $\boldsymbol{\mathcal{Z}} \in \mathbb{C}^{P_\tau \times P_D \times L_\tau \times L_D}$ represents the additive noise tensor, and the factor matrices $\mathbf{B}_{\text{sub},\tau}$, $\mathbf{B}_{\text{sub},D}$, $\mathbf{B}_{\text{shift},\tau}$, and $\mathbf{B}_{\text{shift},D}$ are defined as
\begin{equation}
\begin{aligned}
    \mathbf{B}_{\text{sub},\tau} &= \left[ \mathbf{b}_{\text{sub},\tau}(\tau_1), \dots, \mathbf{b}_{\text{sub},\tau}(\tau_Q) \right] \in \mathbb{C}^{P_\tau \times Q}, \\
    \mathbf{B}_{\text{sub},D} &= \left[ \mathbf{b}_{\text{sub},D}(f_{D,1}), \dots, \mathbf{b}_{\text{sub},D}(f_{D,Q}) \right] \in \mathbb{C}^{P_D \times Q}, \\
    \mathbf{B}_{\text{shift},\tau} &= \left[ \mathbf{b}_{\text{shift},\tau}(\tau_1), \dots, \mathbf{b}_{\text{shift},\tau}(\tau_Q) \right] \in \mathbb{C}^{L_\tau \times Q}, \\
    \mathbf{B}_{\text{shift},D} &= \left[ \mathbf{b}_{\text{shift},D}(f_{D,1}), \dots, \mathbf{b}_{\text{shift},D}(f_{D,Q}) \right] \in \mathbb{C}^{L_D \times Q}.
\end{aligned}
\end{equation}

\begin{figure*}[!t]  
    \centering
    \includegraphics[width=0.60\textwidth]{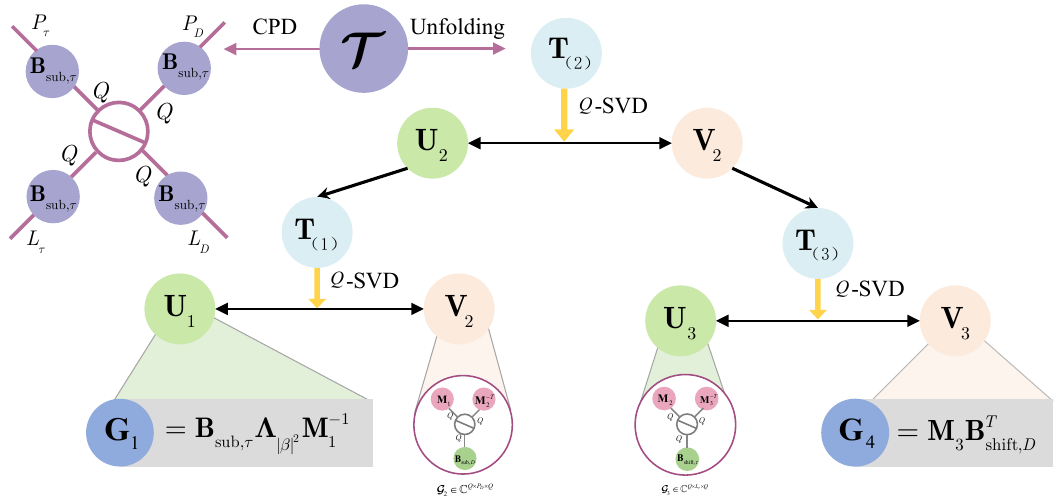} 
    \caption{Binary tree structure of the proposed BT-TTD algorithm. The yellow arrow \protect\YellowArrow\ denotes the Q-SVD operation, which retains only the singular vectors corresponding to the $Q$ largest singular values.  The circles \protect\CircleU\ and \protect\CircleV\ represent the left and right singular vector matrices, respectively, while    \protect\SymbolCPD\  and \protect\SymbolG\ depict the CPD structures of the tensor $\boldsymbol{\mathcal{T}}$ and the TT-cores $\boldsymbol{\mathcal{G}}_2$ and $\boldsymbol{\mathcal{G}}_3$. For more details on graphical representations of tensor networks, see \cite{orus2019tensor}, \cite{evenbly2011tensor} and references therein.}
    \label{fig:bt_tt_process}
\end{figure*}

\section{\textcolor{revisiondarkgreen}{Binary-Tree-Based Tensor Train Decomposition} and Parameter Estimation}

In this section, distinct from the conventional sequential TT-core extraction strategy commonly adopted in \cite{zniyed2019multidimensional,gong2023esprit,oseledets2011tensor}, we propose a novel extraction scheme based on a binary-tree graph topology, as illustrated in Fig. \ref{fig:bt_tt_process}. We theoretically demonstrate that the \textcolor{revisiondarkgreen}{TT-cores} derived via the proposed BT-TTD algorithm are equivalent to the original \textcolor{revisiondarkgreen}{TT-cores}, up to nonsingular transformation matrices. Furthermore, by exploiting the inherent Vandermonde structure within the factor matrices of $\boldsymbol{\mathcal{T}}$, we employ the rectified alternating least squares (RecALS) algorithm to estimate the factor matrices from the extracted \textcolor{revisiondarkgreen}{TT-cores}. Finally, we analyze the computational complexity of the proposed algorithm to highlight its superiority, while the parameter estimation performance is  evaluated in Section~\ref{sec:simulation}.

\subsection{\textcolor{revisiondarkgreen}{Binary-Tree-Based Tensor Train Decomposition (BT-TTD)}}
\label{sec:BT_TTD}

The  workflow of extracting the TT-cores from the tensor 
$\boldsymbol{\mathcal{T}}$ via the proposed BT-TTD algorithm is illustrated in Fig.~\ref{fig:bt_tt_process}. In contrast to the standard sequential extraction methods\cite{oseledets2011tensor}, our approach employs a divide-and-conquer strategy. In this scheme, the tensor indices are recursively split into balanced subsets to construct the tree topology. Moreover, to effectively suppress noise and reduce computational complexity, we adopt the truncated SVD that retains only the singular vectors corresponding to the $Q$ largest singular values, i.e., $Q$-SVD. 

\begin{figure*}[!t]  
    \centering
    \includegraphics[width=0.60\textwidth]{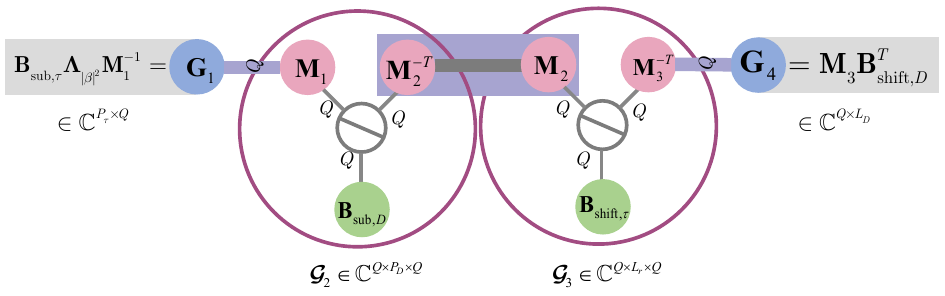} 
    \caption{The analytical structure of the TT-cores. The pink circles \protect\CirclePink\ represent the transformation matrices relative to the original TT-cores, while the green circles \protect\CircleGreen\ denote the factor matrices of the tensor $\boldsymbol{\mathcal{T}}$.}
    \label{fig:Theorem2}
\end{figure*}

In specific, as detailed in the decomposition path, the left singular vector matrix $\mathbf{V}_2 \in \mathbb{C}^{Q \times P_D Q}$ and the right singular vector matrix $\mathbf{U}_3 \in \mathbb{C}^{L_\tau Q \times Q}$ undergo a reshaping operation to reconstruct the  TT-cores $\boldsymbol{\mathcal{G}}_2$ and $\boldsymbol{\mathcal{G}}_3$, respectively. For the boundary nodes, the left singular vector matrix $\mathbf{U}_1$ obtained from $\mathbf{T}_{(1)}$ is directly assigned as the first core $\boldsymbol{\mathcal{G}}_1$ (i.e., $\mathbf{G}_1$), whereas the right singular vector matrix $\mathbf{U}_4$ corresponding to $\mathbf{T}_{(3)}$ is designated as $\boldsymbol{\mathcal{G}}_4$ (i.e., $\mathbf{G}_4$).  Formally, the algebraic structure of the TT-cores recovered by the BT-TTD algorithm is established in the following theorem.

{\itshape \textcolor{revisiondarkgreen}{\textbf{Theorem 2:} The TT-cores
$\boldsymbol{\mathcal{G}}_1$, $\boldsymbol{\mathcal{G}}_2$,
$\boldsymbol{\mathcal{G}}_3$, and $\boldsymbol{\mathcal{G}}_4$
recovered from $\boldsymbol{\mathcal{T}}$ via the BT-TTD algorithm
can be expressed as}}

\begin{equation}  \label{eq:BT_TT_Theorem}
\begin{aligned} \boldsymbol{\mathcal{G}}_1&=\mathbf{G}_1 = \mathbf{B}_{\text{sub},\tau} \mathbf{\Lambda}_{|\beta|^2} \mathbf{M}_1^{-1}, 
\\
\boldsymbol{\mathcal{G}}_2 &= \boldsymbol{\mathcal{I}}_{3,Q} \times_1 \mathbf{M}_1 \times_2 \mathbf{B}_{\text{sub},D} \times_3 \mathbf{M}_2^{-T}, 
 \\ 
\boldsymbol{\mathcal{G}}_3 &= \boldsymbol{\mathcal{I}}_{3,Q} \times_1 \mathbf{M}_2 \times_2 \mathbf{B}_{\text{shift},\tau} \times_3 \mathbf{M}_3^{-T}, 
\\ 
\boldsymbol{\mathcal{G}}_4&=\mathbf{G}_4 = \mathbf{M}_3 \mathbf{B}_{\text{shift},D}^T, \end{aligned}\end{equation}

{\itshape \noindent where \textcolor{revisiondarkgreen}{$\mathbf{\Lambda}_{|\beta|^2} = \operatorname{diag}([|\beta_1|^2, \dots, |\beta_Q|^2])$ denotes the diagonal target-power matrix}, and $\mathbb{M} = \{ \mathbf{M}_k,k=1,2,3 \}$ are \textcolor{revisiondarkgreen}{nonsingular} transformation matrices determined by the SVD truncation process.}
\par\noindent\itshape \textbf{Proof:}  \upshape Please refer to Appendix~\ref{app:proof_theorem_2}. \hfill $\blacksquare$
\par\vspace{\baselineskip}

Theorem 2 indicates that the TT-cores derived via the BT-TTD algorithm differ from the ideal factor cores of $\boldsymbol{\mathcal{T}}$ only by the \textcolor{revisiondarkgreen}{nonsingular} change-of-basis matrices $\mathbb{M}$.  \textcolor{revisiondarkgreen}{This ambiguity is inherent in the TT decomposition} and aligns with the findings in traditional sequential extraction strategies \cite{oseledets2011tensor}. Furthermore, it is evident from \eqref{eq:BT_TT_Theorem} that the desired factor information is  embedded within the structure of the extracted TT-cores $\boldsymbol{\mathcal{G}}_1, \boldsymbol{\mathcal{G}}_2, \boldsymbol{\mathcal{G}}_3, \boldsymbol{\mathcal{G}}_4$, which facilitates the subsequent parameter estimation.



\subsection{Parameter Estimation via Vandermonde Factor Recovery}
\label{sec:param_est}

In this subsection, we proceed to retrieve the  parameters from the TT-cores derived in Section~\ref{sec:BT_TTD}. Recall from Theorem 2 that the factor matrices are preserved within the TT-cores, albeit coupled with nonsingular transformation matrices. Specifically, the information manifests as distinct Vandermonde structures within the column space of $\mathbf{G}_1$, the row space of $\mathbf{G}_4$, and the mode-2 unfoldings of $\boldsymbol{\mathcal{G}}_2$ and $\boldsymbol{\mathcal{G}}_3$. To leverage these structural properties, we employ the \textcolor{revisiondarkgreen}{RecALS algorithm} \cite{zniyed2019multidimensional} to derive the factor matrices $\mathbb B=\{\mathbf{B}_{\text{sub},\tau}, \mathbf{B}_{\text{sub},D},\mathbf{B}_{\text{shift},\tau},\mathbf{B}_{\text{shift},D}\}$. This strategy effectively reconstructs the factors by enforcing the inherent Vandermonde structure during the iterative estimation process.

Based on the derived TT structure, the recovery of $\mathbb{B}$, $\mathbb{M}$
 is achieved by minimizing the following aggregate residual, i.e.,

\begin{equation}
\begin{aligned}
\min_{\mathbb{B}, \mathbb{M}} \: &~ \|\mathbf{G}_1 - \mathbf{B}_{\text{sub},\tau} \mathbf{\Lambda}_{|\beta|^2} \mathbf{M}_1^{-1} \|_F^2 + \|\mathbf{G}_4 - \mathbf{M}_3 \mathbf{B}_{\text{shift},D}^T \|_F^2 \\
& + \|\boldsymbol{\mathcal{G}}_2 - \llbracket \mathbf{M}_1, \mathbf{B}_{\text{sub},D}, \mathbf{M}_2^{-T} \rrbracket \|_F^2 \\
& + \|\boldsymbol{\mathcal{G}}_3 - \llbracket \mathbf{M}_2, \mathbf{B}_{\text{shift},\tau}, \mathbf{M}_3^{-T} \rrbracket \|_F^2.
\end{aligned}
\label{eq:global_cost}
\end{equation}

\noindent It is evident from \eqref{eq:global_cost} that the sub-problems associated with adjacent cores are inextricably linked via the common transformation matrices $\mathbf{M}_k,k=1,2,3$. This coupling renders the simultaneous optimization of all variables computationally prohibitive. Hence, we employ a sequential  strategy, which facilitates the sequential decoupling of the optimization problem.

\subsubsection{RecALS Optimization for $\boldsymbol{\mathcal{G}}_2$}

We initiate the sequential optimization by addressing the \textcolor{revisiondarkgreen}{CPD} of the second core tensor. The objective is to minimize the local fitting error
\begin{equation}
    \min_{\mathbf{M}_1, \mathbf{B}_{\text{sub},D}, \mathbf{M}_2} \left\| \boldsymbol{\mathcal{G}}_2 - \llbracket \mathbf{M}_1, \mathbf{B}_{\text{sub},D}, \mathbf{M}_2^{-T} \rrbracket \right\|_F^2.
\end{equation}
Adopting the canonical alternating least squares (ALS) framework, we update each factor matrix alternately while keeping the others fixed. 

To update the transformation matrix $\mathbf{M}_1$, we first consider the mode-1 unfolding of $\boldsymbol{\mathcal{G}}_2$. Specifically, we have
\begin{equation}
    \mathbf{G}_{2,(1)} = \mathbf{M}_1 \left( \mathbf{M}_2^{-T} \odot \mathbf{B}_{\text{sub},D} \right)^T.
\end{equation}
 Thus, the transformation matrix $\mathbf{M}_1$ is updated as
\begin{equation} \label{eq:update_M1_step}
\begin{aligned}
    \hat{\mathbf{M}}_{1} &= \mathbf{G}_{2,(1)} \left[ \left( \hat{\mathbf{M}}_{2}^{-T} \odot \hat{\mathbf{B}}_{\text{sub},D} \right)^T \right]^{\dagger} \\
    &\overset{(a)}{=} \mathbf{G}_{2,(1)} \left( \hat{\mathbf{M}}_{2}^{-T} \odot \hat{\mathbf{B}}_{\text{sub},D} \right)^{*} \\
    &\quad \times \left[ \left( \hat{\mathbf{M}}_{2}^{-T} \odot \hat{\mathbf{B}}_{\text{sub},D} \right)^T \left( \hat{\mathbf{M}}_{2}^{-T} \odot \hat{\mathbf{B}}_{\text{sub},D} \right)^* \right]^{-1} \\
    &\overset{(b)}{=} \mathbf{G}_{2,(1)} \left( \hat{\mathbf{M}}_{2}^{-T} \odot \hat{\mathbf{B}}_{\text{sub},D} \right)^{*} \\
    &\quad \times \left[ \left( \hat{\mathbf{M}}_{2}^{-1} \hat{\mathbf{M}}_{2}^{-H} \right) \diamond \left( \hat{\mathbf{B}}_{\text{sub},D}^T \hat{\mathbf{B}}_{\text{sub},D}^* \right) \right]^{-1},
\end{aligned}
\end{equation}
where $(a)$ follows from the expansion of the Moore-Penrose pseudoinverse, and $(b)$ leverages the Khatri-Rao product identity $(\mathbf{A} \odot \mathbf{B})^H (\mathbf{A} \odot \mathbf{B}) = (\mathbf{A}^H \mathbf{A}) \diamond (\mathbf{B}^H \mathbf{B})$.

Similarly, to update the factor matrix $\mathbf{B}_{\text{sub},D}$, we  have
\begin{equation} \label{eq:update_B_step}
\begin{aligned}
    \hat{\mathbf{B}}_{\text{sub},D} &= \mathbf{G}_{2,(2)} \left[ \left( \hat{\mathbf{M}}_{2}^{-T} \odot \hat{\mathbf{M}}_{1} \right)^T \right]^{\dagger} \\
    &= \mathbf{G}_{2,(2)} \left( \hat{\mathbf{M}}_{2}^{-T} \odot \hat{\mathbf{M}}_{1} \right)^{*} \\
    &\quad \times \left[ \left( \hat{\mathbf{M}}_{2}^{-1} \hat{\mathbf{M}}_{2}^{-H} \right) \diamond \left( \hat{\mathbf{M}}_{1}^T \hat{\mathbf{M}}_{1}^* \right) \right]^{-1}.
\end{aligned}
\end{equation}
Moreover, since $\mathbf{B}_{\text{sub},D}$ adheres to the Vandermonde structure, its column space exhibits the rotational invariance property, i.e.,
\begin{equation} \label{eq:invariance_B}
    \bar{\mathbf{B}}_{\text{sub},D} = \underline{\mathbf{B}}_{\text{sub},D} \mathbf{Z}_D,
\end{equation}
where $\underline{\mathbf{B}}_{\text{sub},D}\in \mathbb{C}^{P_D-1 \times Q}$ and $\bar{\mathbf{B}}_{\text{sub},D}\in \mathbb{C}^{P_D-1 \times Q}$ are constructed by removing the last and the first rows of $\hat{\mathbf{B}}_{\text{sub},D}$, respectively. Here, $\mathbf{Z}_D = \operatorname{diag}\left( e^{j 2\pi T_{\text{sym}} f_{D,1}}, \dots, e^{j 2\pi T_{\text{sym}} f_{D,Q}} \right)$ denotes the diagonal phase shift matrix containing the Doppler frequency parameters.

Therefore, based on the least squares (LS) principle, the diagonal elements of $\mathbf{Z}_D$, denoted as $\{z_q\}_{q=1}^Q$, are estimated by minimizing the fitting error for each column independently. Assuming the generators lie on the unit circle, we have
\begin{equation} \label{eq:est_z}
\begin{aligned}
    \hat{z}_q &= \arg \min_{|z|=1} \left\| \bar{\mathbf{b}}_q - \underline{\mathbf{b}}_q z \right\|_2^2 \\
    &= \exp\left( j \angle \left( \frac{\underline{\mathbf{b}}_q^H \bar{\mathbf{b}}_q}{\underline{\mathbf{b}}_q^H \underline{\mathbf{b}}_q} \right) \right) = \exp\left( j \angle \left( \underline{\mathbf{b}}_q^H \bar{\mathbf{b}}_q \right) \right),
\end{aligned}
\end{equation}

\noindent where $\underline{\mathbf{b}}_q$ and $\bar{\mathbf{b}}_q$ denote the $q$-th columns of $\underline{\mathbf{B}}_{\text{sub},D}$ and $\bar{\mathbf{B}}_{\text{sub},D}$, respectively. Subsequently, the factor matrix $\mathbf{B}_{\text{sub},D}$ is reconstructed using the estimated generators $\{\hat{z}_q\}_{q=1}^Q$, that is,\begin{equation} \label{eq:reconstruct_B}
    \hat{\mathbf{B}}_{\text{sub},D} = \left[ \mathbf{v}(\hat{z}_1), \mathbf{v}(\hat{z}_2), \dots, \mathbf{v}(\hat{z}_Q) \right] \in \mathbb{C}^{P_D \times Q},
\end{equation}
where $\mathbf{v}(\hat{z}_q) = [1, \hat{z}_q, \hat{z}_q^2, \dots, \hat{z}_q^{P_D-1}]^T$ denotes the reconstructed Vandermonde steering vector determined by the $q$-th generator.

Finally, by leveraging the mode-3 unfolding of $\boldsymbol{\mathcal{G}}_2$, the transformation matrix $\mathbf{M}_2^{-T}$ is updated as
\begin{equation} \label{eq:update_M2_step}
\begin{aligned}
   \hat{\mathbf{M}}_{2}^{-T} &= \mathbf{G}_{2,(3)} \left[ \left(\hat{\mathbf{B}}_{\text{sub},D} \odot \hat{\mathbf{M}}_{1} \right)^T \right]^{\dagger} \\
    &= \mathbf{G}_{2,(3)} \left( \hat{\mathbf{B}}_{\text{sub},D} \odot \hat{\mathbf{M}}_{1} \right)^{*} \\
    &\quad \times \left[ \left(\hat{\mathbf{B}}_{\text{sub},D}^T \hat{\mathbf{B}}_{\text{sub},D}^* \right) \diamond \left( \hat{\mathbf{M}}_{1}^T \hat{\mathbf{M}}_{1}^* \right) \right]^{-1}.
\end{aligned}
\end{equation}
In summary, adopting  the RecALS framework, we iteratively update the factor matrices associated with $\boldsymbol{\mathcal{G}}_2$ in a cyclic manner. Crucially, the Vandermonde manifold projection is integrated into the update step of $\mathbf{B}_{\text{sub},D}$ to strictly enforce the structural constraint. The complete procedure is summarized in Algorithm 1.

\begin{algorithm}
\caption{RecALS-based Factor Estimation for $\boldsymbol{\mathcal{G}}_2$}
\label{alg:recals_G2}
\begin{algorithmic}[1]
    \STATE \textbf{Input:} $\boldsymbol{\mathcal{G}}_2$,  $\epsilon$,  $K_{\max}$.
    \STATE \textbf{Initialization:} Generate random $\hat{\mathbf{B}}_{\text{sub},D}, \hat{\mathbf{M}}_2^{(0)}$; set $k = 0, \Delta = \infty$.

    \WHILE{$k < K_{\max}$ and $\Delta \ge \epsilon$}
        \STATE $k \leftarrow k + 1$

        \STATE Update $\hat{\mathbf{M}}_1^{(k)}$ from  $\mathbf{G}_{2,(1)}$ via \eqref{eq:update_M1_step}.

        \STATE Update $\tilde{\mathbf{B}}$ from $\mathbf{G}_{2,(2)}$ via \eqref{eq:update_B_step}.
        
        \STATE Estimate the generator $\hat{z}_q$ via \eqref{eq:est_z}.
        
        \STATE Reconstruct $\hat{\mathbf{B}}_{\text{sub},D}^{(k)}$ via \eqref{eq:reconstruct_B}.

        \STATE Update $\hat{\mathbf{M}}_2^{-T,(k)}$ from  $\mathbf{G}_{2,(3)}$ via \eqref{eq:update_M2_step}.

        \STATE Calculate relative change $\Delta = ||\hat{\mathbf{B}}_{\text{sub},D}^{(k)} - \hat{\mathbf{B}}_{\text{sub},D}^{(k-1)}||_F / ||\hat{\mathbf{B}}_{\text{sub},D}^{(k-1)}||_F$.
    \ENDWHILE

    \STATE \textbf{Output:} $\hat{\mathbf{M}}_1, \hat{\mathbf{B}}_{\text{sub},D}, \hat{\mathbf{M}}_2^{-T}$. 
\end{algorithmic}
\end{algorithm}

\subsubsection{RecALS Optimization for $\boldsymbol{\mathcal{G}}_3$}

Next, we recover the factors within $\boldsymbol{\mathcal{G}}_3$. Given that $\boldsymbol{\mathcal{G}}_3$ shares the transformation matrix $\mathbf{M}_2$ with $\boldsymbol{\mathcal{G}}_2$, we adopt the previously estimated $\hat{\mathbf{M}}_2$ as a known prior. Consequently, the optimization is simplified to identifying $\mathbf{B}_{\text{shift},\tau}$ and $\mathbf{M}_3$.

First, analogous to \eqref{eq:update_B_step}, the unconstrained update for $\mathbf{B}_{\text{shift},\tau}$ is derived from the mode-2 unfolding $\mathbf{G}_{3,(2)}$ as
\begin{equation} \label{eq:update_B_tau_step}
\begin{aligned}
    \hat{\mathbf{B}}_{\text{shift},\tau} 
    &= \mathbf{G}_{3,(2)} \left( \hat{\mathbf{M}}_{3}^{-T} \odot \hat{\mathbf{M}}_{2} \right)^{*} \\
    &\quad \times \left[ \left( \hat{\mathbf{M}}_{3}^{-1} \hat{\mathbf{M}}_{3}^{-H} \right) \diamond \left( \hat{\mathbf{M}}_{2}^T \hat{\mathbf{M}}_{2}^* \right) \right]^{-1}.
\end{aligned}
\end{equation}
Subsequently, we enforce the Vandermonde manifold constraint similar to the procedure described in \eqref{eq:invariance_B}--\eqref{eq:reconstruct_B}. In specific, the rotational invariance property is exploited to rectify $\hat{\mathbf{B}}_{\text{shift},\tau}$. 

Finally, the transformation matrix $\mathbf{M}_3^{-T}$  is updated as
\begin{equation} \label{eq:update_M3_step}
\begin{aligned}
   \hat{\mathbf{M}}_3^{-T} 
    &= \mathbf{G}_{3,(3)} \left( \hat{\mathbf{B}}_{\text{shift},\tau} \odot \hat{\mathbf{M}}_{2} \right)^{*} \\
    &\quad \times \left[ \left(\hat{\mathbf{B}}_{\text{shift},\tau}^H \hat{\mathbf{B}}_{\text{shift},\tau} \right) \diamond \left( \hat{\mathbf{M}}_{2}^H \hat{\mathbf{M}}_{2} \right) \right]^{-1}.
\end{aligned}
\end{equation}
 Overall, the factor retrieval process for $\boldsymbol{\mathcal{G}}_3$ mirror those of Algorithm 1, excluding the operation in line 5.

 \subsubsection{Recovery of Boundary Factors and Parameter Estimation}
\label{sec:boundary_recovery}

\textcolor{revisiondarkgreen}{Recalling Theorem 2}, the explicit forms of the first and last TT-cores allow us to retrieve $\mathbf{B}_{\text{sub},\tau}$ and $\mathbf{B}_{\text{shift},D}$ via straightforward linear transformations, i.e.,
\begin{equation} \label{eq:recover_boundary}
    \hat{\mathbf{B}}_{\text{sub},\tau} \hat{\mathbf{\Lambda}}_{|\beta|^2} = \mathbf{G}_1 \hat{\mathbf{M}}_1, 
    \hat{\mathbf{B}}_{\text{shift},D} = \mathbf{G}_4^T \hat{\mathbf{M}}_3^{-T}.
\end{equation}

\textcolor{revisiondarkgreen}{Before extracting target-specific physical parameters from the recovered factor matrices, we first establish that the same-indexed columns across these matrices correspond to the same target. This correspondence is formally guaranteed by the following theorem.}

{\color{revisiondarkgreen}
\itshape \textbf{Theorem 3:} Assume that the \textcolor{revisiondarkgreen}{CPDs} of $\boldsymbol{\mathcal{G}}_2$ and $\boldsymbol{\mathcal{G}}_3$ are essentially unique. Under the sequential factor-recovery procedure, there exist a common permutation matrix $\boldsymbol{\Pi}$ and \textcolor{revisiondarkgreen}{nonsingular} diagonal matrices $\boldsymbol{\Delta}_k$, $k=1,\ldots,4$, such that
\begin{equation} \label{eq:common_permutation}
\begin{aligned}
\hat{\mathbf{B}}_{\text{sub},\tau}
&=\mathbf{B}_{\text{sub},\tau}\boldsymbol{\Pi}\boldsymbol{\Delta}_1, &
\hat{\mathbf{B}}_{\text{sub},D}
&=\mathbf{B}_{\text{sub},D}\boldsymbol{\Pi}\boldsymbol{\Delta}_2,\\
\hat{\mathbf{B}}_{\text{shift},\tau}
&=\mathbf{B}_{\text{shift},\tau}\boldsymbol{\Pi}\boldsymbol{\Delta}_3, &
\hat{\mathbf{B}}_{\text{shift},D}
&=\mathbf{B}_{\text{shift},D}\boldsymbol{\Pi}\boldsymbol{\Delta}_4.
\end{aligned}
\end{equation}
\par\noindent\itshape \textbf{Proof:} \upshape Please refer to Appendix~\ref{app:proof_theorem_3}. \hfill $\blacksquare$
}
\par\vspace{\baselineskip}

\textcolor{revisiondarkgreen}{Theorem 3 reveals that all recovered factor matrices share a single arbitrary global permutation $\boldsymbol{\Pi}$, ensuring that same-indexed columns correspond to the same target. This permutation merely reorders the targets jointly, while the scaling ambiguity leaves the column-wise phase structures unchanged.}

Let $\mu_q = -2\pi \Delta f \tau_q$ and $\nu_q = 2\pi T_{\text{sym}} f_{D,q}$ denote the normalized delay and Doppler  frequencies respectively. Then, we have
\begin{equation} \label{eq:est_mu}
    \hat{\mu}_{q} = \frac{ \sum_{\iota_{1}} \angle \left( \frac{ [\hat{\mathbf{b}}_{\text{sub},\tau}(\tau_q)]_{\iota_{1}+1} }{ [\hat{\mathbf{b}}_{\text{sub},\tau}(\tau_q)]_{\iota_{1}} } \right) + \sum_{\iota_{2}} \angle \left( \frac{ [\hat{\mathbf{b}}_{\text{shift},\tau}(\tau_q)]_{\iota_{2}+1} }{ [\hat{\mathbf{b}}_{\text{shift},\tau}(\tau_q)]_{\iota_{2}} } \right) }{ P_{\tau} + L_{\tau} - 2 },
\end{equation}
\begin{equation} \label{eq:est_nu}
    \hat{\nu}_{q} = \frac{ \sum_{\varsigma_{1}} \angle \left( \frac{ [\hat{\mathbf{b}}_{\text{sub},D}(f_{D,q})]_{\varsigma_{1}+1} }{ [\hat{\mathbf{b}}_{\text{sub},D}(f_{D,q})]_{\varsigma_{1}} } \right) + \sum_{\varsigma_{2}} \angle \left( \frac{ [\hat{\mathbf{b}}_{\text{shift},D}(f_{D,q})]_{\varsigma_{2}+1} }{ [\hat{\mathbf{b}}_{\text{shift},D}(f_{D,q})]_{\varsigma_{2}} } \right) }{ P_{D} + L_{D} - 2 },
\end{equation}

\noindent where $\iota_{1} \in\left[1, P_{\tau}-1 \right]$, $\iota_{2} \in\left[1, L_\tau-1 \right]$, $\varsigma_{1} \in\left[1, P_D-1 \right]$ and $\varsigma_{2} \in\left[1, L_{D}-1 \right]$. Here, $\hat{\mathbf{b}}_{\text{sub},\tau}(\tau_q)$ and $\hat{\mathbf{b}}_{\text{shift},\tau}(\tau_q)$ denote the $q$-th columns of $\hat{\mathbf{B}}_{\text{sub},\tau}$ and $\hat{\mathbf{B}}_{\text{shift},\tau}$, respectively; similarly, $\hat{\mathbf{b}}_{\text{sub},D}(f_{D,q})$ and $\hat{\mathbf{b}}_{\text{shift},D}(f_{D,q})$ denote the $q$-th columns of $\hat{\mathbf{B}}_{\text{sub},D}$ and $\hat{\mathbf{B}}_{\text{shift},D}$.

Finally, according to the definitions of $\mu_q$ and $\nu_q$, the  delay and Doppler information can be recovered as
\begin{align}
    \hat{\tau}_q &= -\frac{\hat{\mu}_q}{2\pi \Delta f}, \\
    \hat{f}_{D,q} &= \frac{\hat{\nu}_q}{2\pi T_{\text{sym}}}. \label{eq:final_param}
\end{align}

\noindent This concludes the parameter estimation framework proposed in this work.

\subsection{Complexity Analysis}
The computational complexity of the proposed algorithm comprises three main parts: the 4D tensor construction \eqref{eq:4D_Tensor}, the BT-TTD \eqref{eq:BT_TT_Theorem}, and the RecALS precedures. First, the 4D tensor construction requires $\mathcal{O}(N_{\text{snap}} M_v^2 N_v^2)$ for the autocorrelation matrix computation and $\mathcal{O}(P_\tau P_D L_\tau L_D)$ for tensor reshaping. Second, the BT-TTD complexity is dominated by the $Q$-SVD, requiring $\mathcal{O}(Q \cdot P_\tau P_D L_\tau L_D)+\mathcal{O}(Q^2 (P_\tau P_D + L_\tau L_D))$. Finally, for factor recovery, the intermediate cores, i.e., $\boldsymbol{\mathcal{G}}_2$ and $\boldsymbol{\mathcal{G}}_3$, are iteratively updated using the RecALS algorithm, which incurs a complexity of $\mathcal{O}(I_{\text{iter}} \cdot Q^3(P_D + L_\tau))$ with $I_{\text{iter}}$ being the maximum number of iterations. In contrast, the boundary cores, i.e., $\boldsymbol{\mathcal{G}}_1$ and $\boldsymbol {\mathcal{G}}_4$ are directly recovered via algebraic relations. Overall, the total complexity is  $\mathcal{O}(N_{\text{snap}} M_v^2 N_v^2 + Q P_\tau P_D L_\tau L_D + Q^2 (P_\tau P_D + L_\tau L_D) + I_{\text{iter}} Q^3(P_D + L_\tau))$. For comparative analysis, the complexities of benchmark methods, including CPD-ALS\cite{zhou2017low}, Tensor-ESPRIT\cite{gong2023esprit}, TT-SVD\cite{zniyed2019multidimensional}, 2D-MUSIC\cite{zhang2024ofdm, xie2021performance}, and PhaseUnwrap\cite{wang2025framework, ma2021multi}, are evaluated and summarized in Table \ref{tab:complexity}. The numerical performance  of these benchmark methods will be further investigated in Section \ref{sec:performance}.

\begin{table*}[t]
\centering
\caption{Computational Complexity of Different Algorithms}
\label{tab:complexity}
\renewcommand{\arraystretch}{1.5} 
\begin{tabular}{ c | c | c | c | c }
\hline
\textbf{Algorithms} & \multicolumn{4}{c}{\textbf{Computational Complexity}} \\ \hline

\multirow{2}{*}{\raisebox{-6ex}{Proposed}} & 4D tensor construction $\boldsymbol{\mathcal{T}}$ & BT-TTD & RecALS algorithm & Total Complexity \\ \cline{2-5} 
 & $\mathcal{O}(N_{\text{snap}} M_v^2 N_v^2+P_\tau P_D L_\tau L_D)$ & \parbox{3.5cm}{\centering $\mathcal{O}(Q \cdot P_\tau P_D L_\tau L_D+$ \\$ Q^2 (P_\tau P_D + L_\tau L_D))$ }  & $\mathcal{O}(I_{\text{iter}} \cdot Q^3(P_D + L_\tau))$  & \parbox{4cm}{\centering  $\mathcal{O}(N_{\text{snap}} M_v^2 N_v^2 + Q P_\tau P_D L_\tau L_D +$ \\ $Q^2 (P_\tau P_D + L_\tau L_D) + I_{\text{iter}} Q^3(P_D + L_\tau))$} \\ \hline

2D-MUSIC \cite{zhang2024ofdm, xie2021performance} & \multicolumn{4}{c}{\textcolor{revisiondarkgreen}{$\mathcal{O}\left( N_{\text{snap}} K_{\rm SPB}^2 + K_{\rm SPB}^3 + G_\tau G_D K_{\rm SPB}^2 \right)$}} \\ \hline

PhaseUnwrap \cite{wang2025framework, ma2021multi} & \multicolumn{4}{c}{\textcolor{revisiondarkgreen}{$\mathcal{O} \left( N_\text{snap} \cdot I_\text{iter} Q^2 \left( K_{\rm SPB}^2 + N_\text{RB}M_\text{RB}(G_\tau+G_D) \right) \right)$}} \\ \hline

\multirow{3}{*}{\shortstack{Tensor-based \\ Algorithms}} 
 & Tensor-ESPRIT\cite{gong2023esprit} & \multicolumn{3}{c}{$\mathcal{O}\left( (P_\tau P_D)^2 L_\tau L_D + Q^2 P_\tau P_D + Q^3 + N_{\text{snap}} M_v^2 N_v^2+P_\tau P_D L_\tau L_D\right)$} \\ \cline{2-5}
 & TT-SVD\cite{zniyed2019multidimensional} & \multicolumn{3}{c}{$ \mathcal{O} \left( N_{\text{snap}} M_v^2 N_v^2 + QP_\tau P_D L_\tau L_D + Q^2 ( P_D L_\tau L_D + L_\tau L_D) + I_{\text{iter}} Q^3(P_D + L_\tau)\right) $ }  \\ \cline{2-5}
 & CPD-ALS\cite{zhou2017low},\textcolor{revisiondarkgreen}{\cite{tensorlab3.0}} & \multicolumn{3}{c}{$\mathcal{O} \left( I_{\text{iter}} \cdot \left( P_\tau P_D L_\tau L_D Q + (P_\tau + P_D + L_\tau + L_D)Q^2 + Q^3 \right) \right)$} \\ \hline

\end{tabular}

\vspace{4pt} 
\raggedright 
\footnotesize{\textit{Note:} $I_{\text{iter}}$ denotes the maximum number of iterations. $G_\tau$ and $G_D$ denote the number of search grids in the 2D-MUSIC algorithm.}

\end{table*}

\section{Numerical Results}
\label{sec:simulation}

In this section, representative  simulation results are presented to evaluate the performance of the proposed sensing frame structure and the BT-TTD parameter extraction algorithm. First, the simulation parameter settings are outlined. Then, we compare the parameter estimation performance of the proposed algorithm with that of conventional baseline techniques. Subsequently, the sensing accuracy of different frame structure topologies is evaluated. Finally, we investigate the inherent trade-offs among estimation accuracy, resource occupancy, and computational complexity.

\subsection{Simulation Settings}

\begin{figure}[htbp]
    \centering
    \includegraphics[width=0.85\columnwidth]{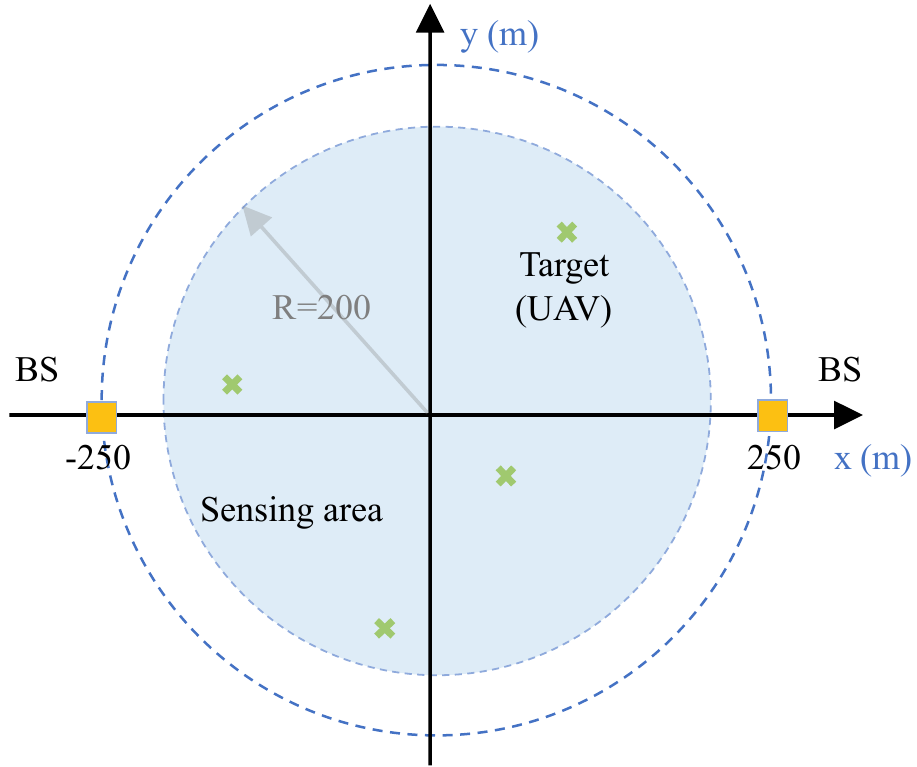}       
    \caption{Top view of the simulation layout, illustrating the locations of BSs and UAV targets.}
    \label{fig:Simulation_Setup}
\end{figure}

For the simulation layout, we consider a bistatic ISAC scenario, as depicted in Fig. \ref{fig:Simulation_Setup}, where the Z-axis representing the altitude is omitted for visual simplicity. Specifically, two single-antenna BSs are respectively deployed at (-250, 0, 30) m and (250, 0, 30) m. The sensing targets are assumed to be three unmanned aerial vehicles (UAVs), i.e., $Q=3$, each with a radar cross-section (RCS) of $\sigma_{\mathrm{RCS}}=0.05\mathrm{ m}^2$. The projections of the UAVs on the X-Y plane are uniformly distributed within a circular area of radius 200 m, and their flight heights are uniformly distributed in $[h_{\min}, h_{\max}]$, where $h_{\min}=$ 20 m and $h_{\max}=$ 40 m. Moreover, the velocity of each UAV is uniformly distributed in $[v_{\min}, v_{\max}]$, where $v_{\min}=$ 2 m/s and $v_{\max}=$ 40 m/s. According to the ITU standard  \cite{ITURRECP525}, the path loss of the bistatic target-scattered paths in dB is defined as\begin{equation}
  \begin{aligned}
    \mathrm{PL}&=103.4+20 \lg (f_{\mathrm{c}}/\text{MHz})+20 \lg (d_{\mathrm{t}}/\text{km}) \\ &\quad +20 \lg (d_{\mathrm{r}}/\text{km})
    -10 \lg (\sigma_{\mathrm{RCS}}/ \mathrm{m}^2)
\end{aligned}  
\end{equation}
\noindent where $f_{\mathrm{c}}$ denotes the central frequency, $d_{\mathrm{t}}$ and $d_{\mathrm{r}}$ denote the distances from the UAV to the Tx and the Rx, respectively. Based on the 5G NR standard \cite{3GPP138104}, we adopt the frequency range 1 (FR1) band and set $f_{\mathrm{c}}$ = 4.9 GHz. For the time-frequency resource allocation, the Tx is configured with a bandwidth part (BWP) of 40 MHz in the frequency domain, comprising an active transmission bandwidth of 38.16 MHz (i.e., 1272 subcarriers) and a 1.84 MHz guard band symmetrically allocated at both ends. {\color{revisiondarkgreen}Meanwhile, the considered frame spans $20\times14=280$ OFDM symbols in time.}\footnote{Unlike \cite{tang2025cooperative, zhang2024target} that dedicate the entire time-frequency grid to sensing, our proposed frame structure sparsely occupies the grid, reserving the remaining resources for communication.} {\color{revisiondarkgreen}Each sensing RB spans $N_{\rm RB}=24$ subcarriers and $M_{\rm RB}=28$ symbols; therefore, the frame contains $F=53$ frequency-block positions, $S=10$ time-block positions, and $N_{\rm snap}=FS=530$ sensing snapshots.} The SCS and the CP duration are set to $\Delta f$ = 30 kHz and $T_{\mathrm{cp}}$ = 2.34 µs, respectively\cite{3GPP13810101}. The noise power spectral density (PSD) at the Rx is set to -174 dBm/Hz.

To evaluate the performance of parameter estimation, the root mean square error (RMSE) \cite{tang2025cooperative} is adopted to quantify the sensing accuracy, i.e., 
\begin{equation}\textcolor{revisiondarkgreen}{\mathrm{RMSE}(\mathbf{x}) = \sqrt{\frac{1}{Q}\sum_{q=1}^{Q}\|\hat{\mathbf{x}}_q - \mathbf{x}_q\|_2^2}},\end{equation}
\noindent where \textcolor{revisiondarkgreen}{$\hat{\mathbf{x}}_q$ and $\mathbf{x}_q$} respectively denote the estimated and the true values of the target parameters \textcolor{revisiondarkgreen}{of the $q$-th target}, including \textcolor{revisiondarkgreen}{its} bistatic range and bistatic relative velocity. 

{\color{revisiondarkgreen}
All algorithms are implemented in MATLAB R2024b and executed on a
64-bit Windows workstation equipped with an AMD Ryzen~7 5800H processor and
16~GB of RAM. The CPD-ALS baseline is implemented using
Tensorlab~\cite{tensorlab3.0} and
initialized by GEVD without random restarts; its relative objective-function
tolerance and maximum number of iterations are set to $10^{-10}$ and 1000,
respectively. For RecALS, $\epsilon$ and $K_{\max}$ in
Algorithm~\ref{alg:recals_G2} are set to $10^{-6}$ and 200, respectively. For
2D-MUSIC, the delay and Doppler grid sizes are set to
$G_\tau=G_D=2000$. At each transmit-power point, the reported result is obtained by
averaging the retained 95\% of the 500 independent Monte Carlo realizations to
mitigate the effect of outliers, following the procedure adopted in
\cite{tang2025cooperative}.
}

\subsection{Parameter Estimation Performance}
\label{sec:performance}

\begin{figure}[!t]
    \centering
    
    \begin{subfigure}[t]{0.85\columnwidth}
        \centering
        \includegraphics[width=\linewidth]{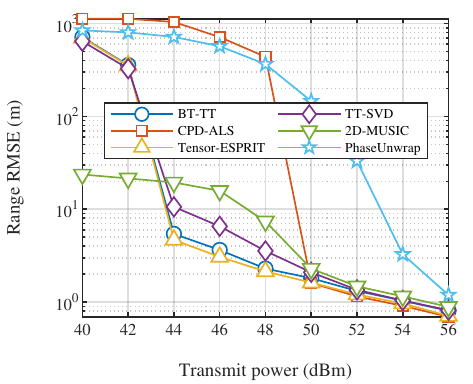}
        \caption{Range estimation RMSE}
        \label{fig:algorithm_range}
    \end{subfigure}
    \par\vspace{0.3\baselineskip}
    \begin{subfigure}[t]{0.85\columnwidth}
        \centering
        \includegraphics[width=\linewidth]{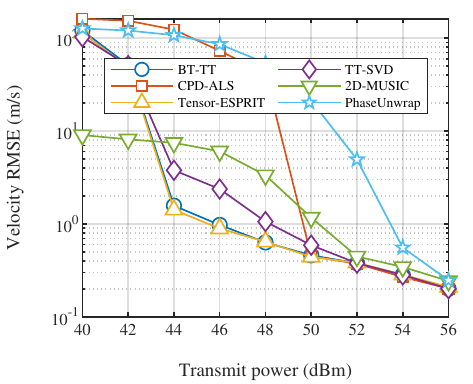}
        \caption{Velocity estimation RMSE}
        \label{fig:algorithm_vel}
    \end{subfigure}
    \par\vspace{0.3\baselineskip}
    \begin{subfigure}[t]{0.85\columnwidth}
        \centering
        \includegraphics[width=\linewidth]{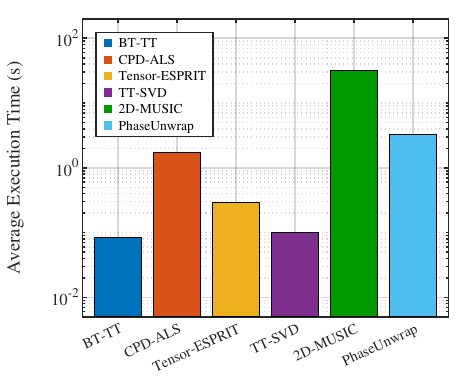}
        \caption{Average CPU execution time}
        \label{fig:CPU_Time}
    \end{subfigure}  
    \caption{Performance comparison of different algorithms vs. the transmit power and the average CPU time.}
    \label{fig:algorithm_performance_comparison}
\end{figure}

Figs. \ref{fig:algorithm_range} and \ref{fig:algorithm_vel}  depict the RMSE of range and velocity estimation versus the transmit power for different algorithms. As expected, the estimation RMSE of all evaluated methods consistently decreases with the increase of transmit power. Notably, tensor-based algorithms, including the proposed BT-TTD, Tensor-ESPRIT, and TT-SVD, exhibit superior estimation accuracy over the entire transmit power range. This performance gain demonstrates the effectiveness of  exploiting the hole-free structural information inherent in the \textcolor{revisiondarkgreen}{autocorrelation matrix}. By comparison, the conventional 2D-MUSIC algorithm fails to exploit this prior knowledge, thus exhibiting a persistent performance gap compared with the tensor-based approaches. Although CPD-ALS is also a tensor-based approach, it fails to utilize the  structural information that the extracted factor matrices are Vandermonde matrices. Consequently, CPD-ALS struggles with convergence issues at low SNRs, leading to a severe performance degradation when the transmit power drops below 48 dBm. Moreover, the PhaseUnwrap method  is observed to be the most sensitive to noise among all the considered baselines, yielding the poorest performance across the transmit power range.

To further evaluate the computational complexity, Fig. \ref{fig:CPU_Time} compares the average CPU execution time. The proposed BT-TTD achieves the highest efficiency, requiring 0.0856 s per estimation, followed by TT-SVD at 0.1021 s. This high efficiency can be attributed to the extraction of  TT-cores, which significantly reduces the dimensionality for the subsequent RecALS. Conversely, Tensor-ESPRIT and CPD-ALS directly operate on the original high-dimensional tensor, requiring 0.2917 s and 1.7072 s, respectively. PhaseUnwrap requires 3.2646 s, whereas 2D-MUSIC incurs the largest runtime of 31.6755 s. In summary, BT-TTD delivers near-optimal estimation performance with the lowest computational overhead, achieving a highly favorable trade-off.

\subsection{\textcolor{revisiondarkgreen}{Performance Comparison of Different Sensing Frame Structures}}

\begin{table}[!t]
\centering
\captionsetup{labelfont={color=revisiondarkgreen},textfont={color=revisiondarkgreen}}
\caption{Comparison of contiguous virtual URA sizes and critical lag-redundancy weights.}
\label{tab:udofs_comparison}
{\color{revisiondarkgreen}
\setlength{\tabcolsep}{4pt}
\begin{tabular}{c c c c c c}
\hline
Array structure & $M_v\times N_v$ & $w(1,0)$ & $w(0,1)$ & $w(1,1)$ & $w(1,-1)$ \\
\hline
HA       & $47\times55$ & 10 &  2 &  7 &  7 \\
CPA      & $18\times22$ &  2 &  2 &  2 &  2 \\
HOBA     & $47\times55$ & 46 &  2 &  1 &  1 \\
OBA      & $47\times55$ & 46 & 27 &  1 &  1 \\
HOBA-2   & $47\times55$ &  6 &  2 & 21 & 21 \\
SSCPA    & $43\times43$ &  2 &  2 &  1 &  1 \\
LA       & $25\times55$ &  2 &  3 & 13 & 13 \\
Thermos  & $47\times49$ &  3 &  3 &  7 &  6 \\
HHA      & $25\times55$ &  2 & 27 &  1 &  1 \\
\hline
\end{tabular}
}
\end{table}

{\color{revisiondarkgreen}
As shown in Section~\ref{sec:virtual URA Construction}, constructing the 4D tensor $\boldsymbol{\mathcal T}$ requires only a contiguous virtual URA $\mathbb U\subseteq\mathbb D_{\rm SPB}$, rather than an entirely hole-free difference coarray. Motivated by this property, we quantitatively compare nine representative 2D sparse geometries under the same OFDM/ISAC configuration: HA, CPA, HOBA, OBA, HOBA-2, SSCPA, LA, Thermos, and HHA \cite{liu2017hourglass,vaidyanathan2010sparse,ren2018extensions,zhang2025shift,sun2018thermos}. Table~\ref{tab:udofs_comparison} summarizes the contiguous virtual URA size and critical lag-redundancy weights of each structure.
}

\begin{figure}[!t]
    \centering
    \begin{subfigure}[t]{0.85\columnwidth}
        \centering
        \includegraphics[width=\linewidth]{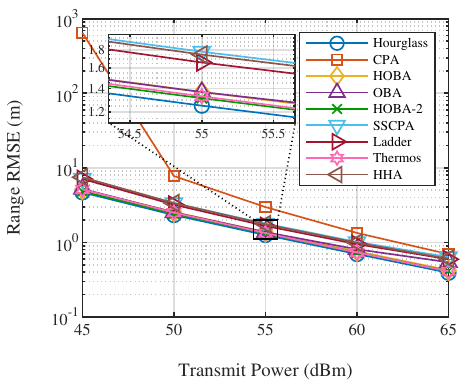}
        \caption{Range estimation RMSE}
        \label{fig:range}
    \end{subfigure}

    \vspace{0.4\baselineskip}

    \begin{subfigure}[t]{0.85\columnwidth}
        \centering
        \includegraphics[width=\linewidth]{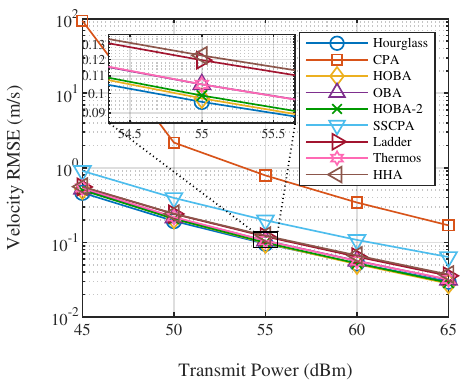}
        \caption{Velocity estimation RMSE}
        \label{fig:vel}
    \end{subfigure}
    \captionsetup{labelfont={color=revisiondarkgreen},textfont={color=revisiondarkgreen}}
    \caption{RMSE performance of range and velocity estimation versus the transmit power for different sensing frame structures.}
    \label{fig:Different_Array_rmse}
\end{figure}

{\color{revisiondarkgreen}
Fig.~\ref{fig:Different_Array_rmse} compares the range and velocity estimation RMSEs of the nine sensing frame structures over the considered transmit-power range. The RMSEs of all structures decrease as the transmit power increases. When interpreted together with Table~\ref{tab:udofs_comparison}, the results reveal a broad aperture-related trend. HA, HOBA, OBA, and HOBA-2 each provide the largest contiguous virtual URA of $47\times55$, whereas Thermos provides a $47\times49$ virtual URA and retains the full frequency-domain aperture; together, these structures constitute the leading performance group. By contrast, CPA, with the smallest virtual URA of $18\times22$, exhibits the largest RMSEs, whereas SSCPA, LA, and HHA show intermediate performance. However, HA, HOBA, OBA, and HOBA-2 still exhibit different RMSEs despite sharing the same $47\times55$ virtual URA, indicating that aperture size alone cannot fully account for their performance differences.

The lag-redundancy profile provides a complementary explanation. Since the entry-wise noise variance in $\mathbf U_{\rm vir}$ is inversely related to $w(p,q)$, a more balanced redundancy distribution reduces noise nonuniformity across the virtual array. HA therefore remains in the leading group and provides the best overall range--velocity trade-off, benefiting jointly from its large contiguous virtual aperture and relatively balanced lag-redundancy distribution.
}

\begin{figure}[htbp]
    \centering
    \begin{subfigure}[t]{0.49\linewidth}
        \centering
        \includegraphics[width=\linewidth]{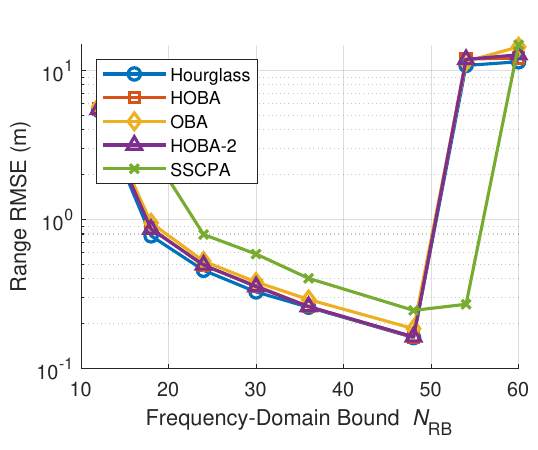} 
        \caption{Range RMSE}
        \label{fig:Tradeoff_sub_range}
    \end{subfigure}
    \hfill 
    \begin{subfigure}[t]{0.49\linewidth}
        \centering
        \includegraphics[width=\linewidth]{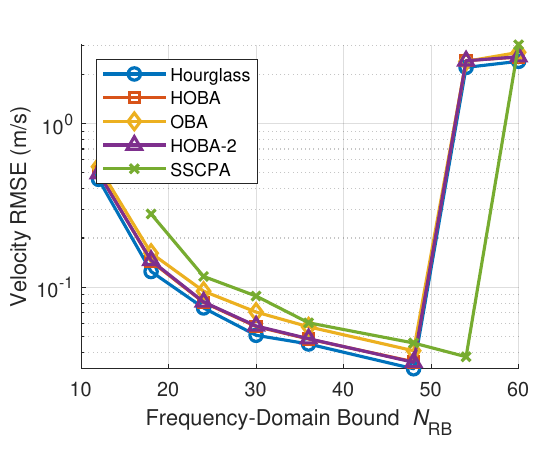}
        \caption{Velocity RMSE}
        \label{fig:Tradeoff_sub_velocity}
    \end{subfigure}
    
    \vspace{0.3\baselineskip} 
    
    \begin{subfigure}[t]{0.49\linewidth}
        \centering
        \includegraphics[width=\linewidth]{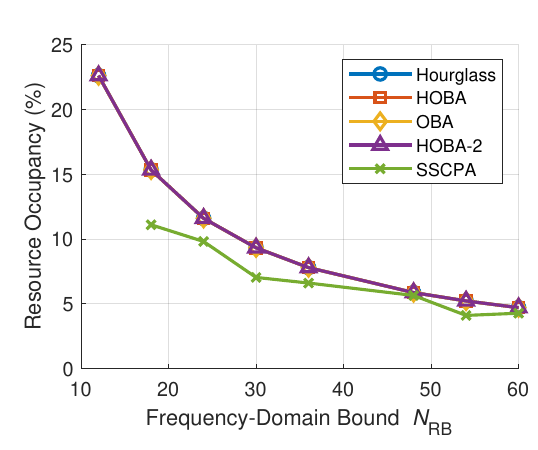}
        \caption{Resource Occupancy}
        \label{fig:Tradeoff_sub_overhead}
    \end{subfigure}
    \hfill
    \begin{subfigure}[t]{0.49\linewidth}
        \centering
        \includegraphics[width=\linewidth]{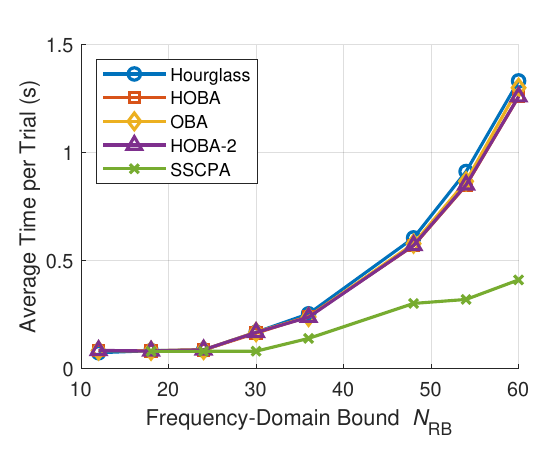}
        \caption{Computational Complexity}
        \label{fig:Tradeoff_sub_computation}
    \end{subfigure}
    
    \caption{Performance trade-offs of different array topologies vs. the  time-frequency pattern dimensions. The RB dimensions are scaled at a fixed ratio of $N_\text{RB}:M_\text{RB} = 6:7$.}
    \label{fig:overall_tradeoff}
\end{figure}

\subsection{Trade-offs among Estimation Accuracy, Resource Occupancy, and Computational Complexity}
\label{sec:tradeoff}
 
 Fig. \ref{fig:overall_tradeoff} evaluates the trade-offs among sensing accuracy, resource occupancy, and computational complexity versus pattern dimensions at 50 dBm transmit power. As Figs. \ref{fig:Tradeoff_sub_range} and \ref{fig:Tradeoff_sub_velocity} depict, the estimation RMSE initially drops with a larger $N_{\text{RB}}$.  However, for $N_{\text{RB}} > 48$, performance sharply degrades, driven by diminished smoothing snapshots from excessive pattern expansion within the fixed time-frequency grid.  Consequently, the \textcolor{revisiondarkgreen}{autocorrelation matrix} becomes rank-deficient and ill-conditioned, resulting in parameter extraction failure. Moreover, Figs. \ref{fig:Tradeoff_sub_overhead} and \ref{fig:Tradeoff_sub_computation} confirm that while enlarging $N_{\text{RB}}$ up to 48 improves precision and reduces resource occupancy, it incurs  computational overhead. Therefore, the pattern dimension requires careful calibration to balance accuracy, efficiency, and complexity.
 
\section{Conclusion}
This paper has presented a  RB-level sparse sensing framework for 5G NR ISAC systems. The proposed design, integrated with the novel BT-TTD and Vandermonde-prior-guided RecALS algorithms, offers a highly efficient approach for  parameter estimation. Numerical simulations demonstrate the near-optimal estimation accuracy and drastically reduced computational burden of our scheme. Furthermore, our topological analysis reveals the inherent trade-offs, paving the way for practical OFDM-ISAC implementations in the future wireless networks. Future work will explore extending this framework to networked ISAC with multi-BS cooperation and addressing practical deployment challenges, such as interference cancellation in high-mobility scenarios.

\appendices
\section{Proof of Theorem 1 \label{app:proof_theorem_1}}

{\color{revisiondarkgreen}
In this appendix, we derive the exact CRB for joint delay and Doppler estimation stated in Theorem 1. For the single-target case, i.e., $Q=1$, the target subscript $q$ is omitted for brevity. We first vectorize the sensing channel $\mathbf{H}_r$ as
\begin{equation}
\begin{aligned}
\mathbf{y}
&=\operatorname{vec}(\mathbf{H}_r)\\
&=\operatorname{diag}\!\left(\operatorname{vec}(\mathbf{U}_r)\right)
\left(
\beta\mathbf{a}_D^{*}(f_D)
\otimes
\mathbf{a}_{\tau}(\tau)
\right)
+\operatorname{vec}(\tilde{\mathbf{Z}})\\
&=\beta\mathbf{\Lambda}\mathbf{v}(\boldsymbol{\xi})
+\mathbf{z}
\in\mathbb{C}^{MN\times1},
\end{aligned}
\end{equation}
where $\mathbf{\Lambda}=\operatorname{diag}(\operatorname{vec}(\mathbf{U}_r))$ is the binary selection diagonal matrix, $\mathbf{v}(\boldsymbol{\xi})=\mathbf{a}_D^{*}(f_D)\otimes\mathbf{a}_{\tau}(\tau)$ is the full-dimensional steering vector, $\boldsymbol{\xi}=[\tau,f_D]^T$, and $\mathbf{z}\sim\mathcal{CN}(\mathbf{0},\sigma_z^2\mathbf{I}_{MN})$.

Then, by extracting the $K_{\rm frame}$ entries indexed by $\mathbb S_{\rm frame}$, we obtain the effective observation vector as
\begin{equation}
\tilde{\mathbf{y}}
=
\beta\tilde{\mathbf{a}}(\boldsymbol{\xi})
+\tilde{\mathbf{z}}
\in\mathbb{C}^{K_{\rm frame}\times1},
\end{equation}
where $\tilde{\mathbf{z}}\sim\mathcal{CN}(\mathbf{0},\sigma_z^2\mathbf{I}_{K_{\rm frame}})$, and the $k$-th entry of the effective steering vector is
\[
[\tilde{\mathbf{a}}(\boldsymbol{\xi})]_k
=
e^{-j2\pi n_k\Delta f\tau}
e^{j2\pi m_kT_{\mathrm{sym}}f_D},
\qquad k=1,\ldots,K_{\rm frame}.
\]

By treating the unknown complex coefficient $\beta$ as a nuisance parameter, the CRB matrix for $\boldsymbol{\xi}=[\tau,f_D]^T$ can be obtained as \cite{hu2024joint}
\begin{equation}
\operatorname{CRB}(\boldsymbol{\xi})
=
\frac{\sigma_z^2}{2|\beta|^2}\mathbf{J}^{-1},
\label{eq:AppA_CRB}
\end{equation}
where $\mathbf{J}$ is the normalized information matrix given by
\begin{equation}
\mathbf{J}
=
\operatorname{Re}
\left\{
\mathbf{D}_{\boldsymbol{\xi}}^H
\mathbf{P}_{\tilde{\mathbf{a}}}^{\perp}
\mathbf{D}_{\boldsymbol{\xi}}
\right\}
\in\mathbb{R}^{2\times2}.
\end{equation}
The derivative matrix $\mathbf{D}_{\boldsymbol{\xi}}$ collects the sensitivities of the effective steering vector to delay and Doppler, i.e.,
\begin{equation}
\mathbf{D}_{\boldsymbol{\xi}}
=
\left[
\mathbf{d}_{\tau},
\mathbf{d}_{f_D}
\right]
=
\left[
\frac{\partial\tilde{\mathbf{a}}}{\partial\tau},
\frac{\partial\tilde{\mathbf{a}}}{\partial f_D}
\right].
\end{equation}
where $\mathbf{d}_{\tau}=-j2\pi\Delta f(\mathbf{n}\diamond\tilde{\mathbf{a}})$ and $\mathbf{d}_{f_D}=j2\pi T_{\mathrm{sym}}(\mathbf{m}\diamond\tilde{\mathbf{a}})$. Here, $\mathbf{n}=[n_1,\ldots,n_{K_{\rm frame}}]^T$ and $\mathbf{m}=[m_1,\ldots,m_{K_{\rm frame}}]^T$ denote the frame-level subcarrier and OFDM-symbol positions, respectively.

To simplify the derivation of $\mathbf{J}$, let $\bar n=K_{\rm frame}^{-1}\sum_{k=1}^{K_{\rm frame}}n_k$ and $\bar m=K_{\rm frame}^{-1}\sum_{k=1}^{K_{\rm frame}}m_k$ denote the mean subcarrier and OFDM-symbol positions, respectively, and define the corresponding population variances and covariance as
\begin{equation}
\begin{aligned}
V_n
&=
\frac{1}{K_{\rm frame}}\sum_{k=1}^{K_{\rm frame}}(n_k-\bar n)^2,\\
V_m
&=
\frac{1}{K_{\rm frame}}\sum_{k=1}^{K_{\rm frame}}(m_k-\bar m)^2,\\
C_{nm}
&=
\frac{1}{K_{\rm frame}}\sum_{k=1}^{K_{\rm frame}}
(n_k-\bar n)(m_k-\bar m).
\end{aligned}
\end{equation}

\noindent Accordingly, the delay-related diagonal entry of $\mathbf{J}$ is
\begin{equation}
\begin{aligned}
[\mathbf{J}]_{1,1}
&=
\mathbf{d}_{\tau}^H
\mathbf{P}_{\tilde{\mathbf{a}}}^{\perp}
\mathbf{d}_{\tau}\\
&=
4\pi^2\Delta f^2
\sum_{k=1}^{K_{\rm frame}}(n_k-\bar n)^2\\
&=
4\pi^2K_{\rm frame}\Delta f^2V_n.
\end{aligned}
\label{eq:AppA_J11}
\end{equation}
Similarly, the Doppler-related diagonal element is
\begin{equation}
[\mathbf{J}]_{2,2}=4\pi^2K_{\rm frame}T_{\mathrm{sym}}^2V_m.
\label{eq:AppA_J22}
\end{equation}
Next, using the definitions of $\bar n$, $\bar m$, and $C_{nm}$, we express the off-diagonal entry for delay--Doppler coupling as
\begin{equation}
\begin{aligned}
[\mathbf{J}]_{1,2}
={}&
\operatorname{Re}
\left\{
\mathbf{d}_{\tau}^H
\mathbf{P}_{\tilde{\mathbf{a}}}^{\perp}
\mathbf{d}_{f_D}
\right\}\\
={}&
-4\pi^2\Delta fT_{\mathrm{sym}}
\sum_{k=1}^{K_{\rm frame}}
(n_k-\bar n)(m_k-\bar m)\\
={}&
-4\pi^2K_{\rm frame}\Delta fT_{\mathrm{sym}}C_{nm}.
\end{aligned}
\label{eq:AppA_J12}
\end{equation}
Since $\mathbf{J}$ is real and symmetric, $[\mathbf{J}]_{2,1}=[\mathbf{J}]_{1,2}$. The complete core information matrix therefore takes the form
\begin{equation}
\mathbf{J}
=
4\pi^2K_{\rm frame}
\begin{bmatrix}
\Delta f^2V_n
&
-\Delta fT_{\mathrm{sym}}C_{nm}
\\
-\Delta fT_{\mathrm{sym}}C_{nm}
&
T_{\mathrm{sym}}^2V_m
\end{bmatrix}.
\label{eq:AppA_J_exact}
\end{equation}

The conditions $V_n>0$, $V_m>0$, and $|C_{nm}|<\sqrt{V_nV_m}$ ensure that $\mathbf{J}$ is invertible. Therefore, the exact CRB matrix takes the form
\begin{equation}
\operatorname{CRB}(\boldsymbol{\xi})
=
\kappa_{\rho}
\begin{bmatrix}
\dfrac{1}{\Delta f^2V_n}
&
\dfrac{\rho}{\Delta fT_{\mathrm{sym}}\sqrt{V_nV_m}}
\\[2mm]
\dfrac{\rho}{\Delta fT_{\mathrm{sym}}\sqrt{V_nV_m}}
&
\dfrac{1}{T_{\mathrm{sym}}^2V_m}
\end{bmatrix},
\end{equation}
where
\begin{equation}
\kappa_{\rho}
=
\frac{\sigma_z^2}
{8\pi^2|\beta|^2K_{\rm frame}(1-\rho^2)},
\qquad
\rho
=
\frac{C_{nm}}{\sqrt{V_nV_m}}.
\end{equation}
Accordingly, the marginal CRBs are
\begin{subequations}
\begin{align}
\operatorname{CRB}(\tau)
&=
\frac{\sigma_z^2}
{8\pi^2|\beta|^2K_{\rm frame}\Delta f^2V_n(1-\rho^2)},\\
\operatorname{CRB}(f_D)
&=
\frac{\sigma_z^2}
{8\pi^2|\beta|^2K_{\rm frame}T_{\mathrm{sym}}^2V_m(1-\rho^2)}.
\end{align}
\end{subequations}

\noindent This concludes the proof. \hfill $\blacksquare$
}

\section{\textcolor{revisiondarkgreen}{Proof of Proposition 1} \label{app:proof_proposition_1}}

{\color{revisiondarkgreen}
From \eqref{eq:beta_block} and \eqref{eq:R_beta_RB}, the $(q,\ell)$-th entry of $\mathbf R_{\beta,{\rm RB}}$ is
\begin{equation}
\begin{aligned}
\left[\mathbf R_{\beta,{\rm RB}}\right]_{q,\ell}
&=\frac{1}{FS}\sum_{f=0}^{F-1}\sum_{s=0}^{S-1}
\beta_{q,f,s}\beta_{\ell,f,s}^{*}\\
&=\beta_q\beta_\ell^{*}
D_F\!\left(-2\pi N_{\rm RB}\Delta f\Delta\tau_{q,\ell}\right)\\
&\quad\times
D_S\!\left(2\pi M_{\rm RB}T_{\rm sym}\Delta f_{D,q,\ell}\right),
\end{aligned}
\end{equation}
where $\Delta\tau_{q,\ell}=\tau_q-\tau_\ell$ and $\Delta f_{D,q,\ell}=f_{D,q}-f_{D,\ell}$. For $q=\ell$, both differences are zero and $D_F(0)=D_S(0)=1$; hence, $[\mathbf R_{\beta,{\rm RB}}]_{q,q}=|\beta_q|^2$. For $q\neq\ell$, the magnitude of the corresponding off-diagonal entry satisfies
\begin{equation}
\begin{aligned}
\left|\left[\mathbf R_{\beta,{\rm RB}}\right]_{q,\ell}\right|
&=|\beta_q||\beta_\ell|
\left|
D_F\!\left(2\pi N_{\rm RB}\Delta f\Delta\tau_{q,\ell}\right)
\right|\\
&\quad\times
\left|
D_S\!\left(2\pi M_{\rm RB}T_{\rm sym}\Delta f_{D,q,\ell}\right)
\right|\\
&\leq\epsilon|\beta_q||\beta_\ell|,
\end{aligned}
\end{equation}
where $|D_F(-x)|=|D_F(x)|$ has been used. The first nonzero nulls of the delay- and Doppler-domain Dirichlet kernels determine the resolutions
\begin{equation}\label{eq:first_null_resolutions}
\begin{aligned}
\Delta\tau_0
&=\frac{1}{FN_{\rm RB}\Delta f}
=\frac{1}{N\Delta f},\\
\Delta f_{D,0}
&=\frac{1}{SM_{\rm RB}T_{\rm sym}}
=\frac{1}{MT_{\rm sym}}.
\end{aligned}
\end{equation}
Since the Dirichlet kernels are periodic, we define the minimum periodic delay and Doppler separations as
\begin{equation}\label{eq:principal_separations}
\begin{aligned}
\widetilde{\Delta\tau}_{q,\ell}
&\triangleq
\min_{k_\tau\in\mathbb Z}
\left|
\Delta\tau_{q,\ell}
-\frac{k_\tau}{N_{\rm RB}\Delta f}
\right|,\\
\widetilde{\Delta f}_{D,q,\ell}
&\triangleq
\min_{k_D\in\mathbb Z}
\left|
\Delta f_{D,q,\ell}
-\frac{k_D}{M_{\rm RB}T_{\rm sym}}
\right|,
\end{aligned}
\end{equation}
where $k_\tau,k_D\in\mathbb Z$ denote the delay- and Doppler-domain ambiguity indices, respectively. By construction, $0\leq\widetilde{\Delta\tau}_{q,\ell}\leq\Delta\tau_{\rm p}$ and $0\leq\widetilde{\Delta f}_{D,q,\ell}\leq\Delta f_{D,{\rm p}}$, where $\Delta\tau_{\rm p}\triangleq1/(2N_{\rm RB}\Delta f)$ and $\Delta f_{D,{\rm p}}\triangleq1/(2M_{\rm RB}T_{\rm sym})$ are the principal-separation limits. Accordingly, the maximum sidelobe magnitudes are defined as
\begin{equation}\label{eq:dirichlet_sidelobe_bound}
\begin{aligned}
\rho_F
&\triangleq
\max_{\Delta\tau_0\leq\widetilde{\Delta\tau}\leq\Delta\tau_{\rm p}}
\left|D_F\!\left(2\pi N_{\rm RB}\Delta f\widetilde{\Delta\tau}\right)\right|,\\
\rho_S
&\triangleq
\max_{\Delta f_{D,0}\leq\widetilde{\Delta f}_D\leq\Delta f_{D,{\rm p}}}
\left|D_S\!\left(2\pi M_{\rm RB}T_{\rm sym}\widetilde{\Delta f}_D\right)\right|.
\end{aligned}
\end{equation}
If, for every $q\neq\ell$, the principal separations satisfy
\begin{equation}\label{eq:resolution_condition}
\begin{aligned}
\widetilde{\Delta\tau}_{q,\ell}
&\geq\Delta\tau_0,\\
\widetilde{\Delta f}_{D,q,\ell}
&\geq\Delta f_{D,0},
\end{aligned}
\end{equation}
then, by periodicity and even symmetry, the corresponding kernel product satisfies
\begin{equation}\label{eq:dirichlet_product_bound}
\begin{aligned}
&\left|D_F\!\left(2\pi N_{\rm RB}\Delta f\widetilde{\Delta\tau}_{q,\ell}\right)\right|\\[-0.2em]
{}\times{}&
\left|D_S\!\left(2\pi M_{\rm RB}T_{\rm sym}\widetilde{\Delta f}_{D,q,\ell}\right)\right|
\leq \rho_F\rho_S.
\end{aligned}
\end{equation}
Under the simulation settings in Section~VI-A, $F=53$ and $S=10$. Equation~\eqref{eq:dirichlet_sidelobe_bound} gives $\rho_F\approx0.2175$ and $\rho_S\approx0.2247$, and hence
\begin{equation}\label{eq:numerical_sidelobe_bound}
\rho_F\rho_S\approx0.0489<0.05.
\end{equation}
Therefore, when \eqref{eq:resolution_condition} holds for all target pairs, $\epsilon\leq\rho_F\rho_S<0.05$, and \eqref{eq:R_beta_error_bound} implies $\mathbf R_{\beta,{\rm RB}}\approx\mathbf R_s$. This concludes the proof. \hfill $\blacksquare$
}

\section{Proof of Theorem 2 \label{app:proof_theorem_2}}

In this appendix, we verify the relationship between the core tensor derived from the BT-TTD decomposition and the original core tensor. To facilitate the derivation, we introduce the following \textcolor{revisiondarkgreen}{lemma}.

\textit{\textbf{Lemma 1:} (Subspace equivalence) \textcolor{revisiondarkgreen}{\cite{golub2013matrix}}}
Let $\mathbf{X} = \mathbf{L}\mathbf{R} \in \mathbb{C}^{M \times N}$ denote a rank-$K$ matrix, where $\mathbf{L} \in \mathbb{C}^{M \times K}$ and $\mathbf{R} \in \mathbb{C}^{K \times N}$ are full-rank factor matrices. Let $\mathbf{X} = \mathbf{U}_K \mathbf{\Sigma}_K \mathbf{V}_K^H$ denote its compact SVD. Then, the column and row spaces satisfy
\begin{equation}
    \text{range}(\mathbf{U}_K) = \text{range}(\mathbf{L}), \quad \text{range}(\mathbf{V}_K) = \text{range}(\mathbf{R}^H).
\end{equation}
Consequently, there exist \textcolor{revisiondarkgreen}{nonsingular} transformation matrices $\mathbf{T}_L, \mathbf{T}_R \in \mathbb{C}^{K \times K}$ such that
\begin{equation}
    \mathbf{U}_K = \mathbf{L}\mathbf{T}_L, \quad \mathbf{V}_K = \mathbf{R}^H \mathbf{T}_R.
\end{equation}

\vspace{2pt} 


\noindent\textcolor{revisiondarkgreen}{\textit{\textbf{Proof of Theorem 2:}}}
Based on the BT-TTD splitting mechanism, the \textcolor{revisiondarkgreen}{mode-2 unfolding} of $\boldsymbol{\mathcal{T}}$ is given by
\begin{equation}
    \mathbf{T}_{(2)} = (\mathbf{B}_{\text{sub},D} \odot \mathbf{B}_{\text{sub},\tau}) \mathbf{\Lambda}_{|\beta|^2} (\mathbf{B}_{\text{shift},D} \odot \mathbf{B}_{\text{shift},\tau})^T.
\end{equation}

\noindent Then, let $\mathbf{U}_2$ denote the left singular matrix obtained from the compact SVD of $\mathbf{T}_{(2)}$. According to the subspace equivalence established in \textit{\textbf{Lemma 1}}, we have
\begin{equation}
    \mathbf{U}_2 = (\mathbf{B}_{\text{sub},D} \odot \mathbf{B}_{\text{sub},\tau}) \mathbf{\Lambda}_{|\beta|^2} \mathbf{M}_2^{-1},
\end{equation}
where $\mathbf{M}_2$ is a \textcolor{revisiondarkgreen}{nonsingular} basis transformation matrix. \textcolor{revisiondarkgreen}{It is readily verified that $\mathbf{U}_2^T$ is the mode-1 unfolding of $\boldsymbol{\mathcal{W}}$, i.e., $\mathbf{U}_2^T=\mathbf{W}_{(1)}$, where}
\begin{equation}
\textcolor{revisiondarkgreen}{\boldsymbol{\mathcal{W}} = \boldsymbol{\mathcal{I}}_{3,Q} \times_1 (\mathbf{M}_2^{-T}\mathbf{\Lambda}_{|\beta|^2}) \times_2 \mathbf{B}_{\text{sub},\tau} \times_3 \mathbf{B}_{\text{sub},D}.}
\end{equation}
\textcolor{revisiondarkgreen}{Hence, by rearranging the matrix elements, we define $\mathbf{T}_{(1)}$ as the mode-2 unfolding of $\boldsymbol{\mathcal{W}}$, i.e.,}
\begin{equation}
 \textcolor{revisiondarkgreen}{\mathbf{T}_{(1)} \triangleq \mathbf{W}_{(2)}} = \mathbf{B}_{\text{sub},\tau} \mathbf{\Lambda}_{|\beta|^2}(\mathbf{B}_{\text{sub},D} \odot \mathbf{M}_2^{-T})^T.
\end{equation}
\noindent Next, performing a compact SVD on $\mathbf{T}_{(1)}$ gives $\mathbf{T}_{(1)} = \mathbf{U}_1 \mathbf{V}_1$, where the singular values are absorbed into the left factor $\mathbf{U}_1\in \mathbb{C}^{P_\tau\times Q}$.  Due to the property of   \textit{\textbf{Lemma 1}}, we have
\begin{equation}\mathbf{U}_1 = \mathbf{B}_{\text{sub},\tau} \mathbf{\Lambda}_{|\beta|^2} \mathbf{M}_1^{-1},
\end{equation}
where $\mathbf{M}_1^{-1} \in \mathbb{C}^{Q \times Q}$ is the \textcolor{revisiondarkgreen}{nonsingular} basis transformation matrix. By definition, $\mathbf{U}_1$ directly constitutes the first TT-core $\boldsymbol{\mathcal{G}}_1$, i.e.,
\begin{equation}
 \boldsymbol{\mathcal{G}}_1 = \mathbf{B}_{\text{sub},\tau} \mathbf{\Lambda}_{|\beta|^2} \mathbf{M}_1^{-1}.\end{equation}
\noindent To derive $\boldsymbol{\mathcal{G}}_2$, we substitute $\mathbf{U}_1$ back into the SVD equivalence $\mathbf{U}_1 \mathbf{V}_1 = \mathbf{T}_{(1)}$, which  yields
 \begin{equation} (\mathbf{B}_{\text{sub},\tau} \mathbf{\Lambda}_{|\beta|^2} \mathbf{M}_1^{-1}) \mathbf{V}_1 = \mathbf{B}_{\text{sub},\tau} \mathbf{\Lambda}_{|\beta|^2} (\mathbf{B}_{\text{sub},D} \odot \mathbf{M}_2^{-T})^T.
\end{equation}
Since the left factor $\mathbf{B}_{\text{sub},\tau} \mathbf{\Lambda}_{|\beta|^2}$ has full column rank, we left-multiply both sides by its \textcolor{revisiondarkgreen}{Moore--Penrose pseudoinverse} $\mathbf{L}^\dagger = (\mathbf{L}^H \mathbf{L})^{-1} \mathbf{L}^H$ to obtain
\begin{equation}\mathbf{V}_1 = \mathbf{M}_1 (\mathbf{B}_{\text{sub},D} \odot \mathbf{M}_2^{-T})^T,\end{equation}
where $\mathbf{L} = \mathbf{B}_{\text{sub},\tau} \mathbf{\Lambda}_{|\beta|^2}\in \mathbb{C}^{P_\tau\times Q}$. Recognizing that $\mathbf{V}_1$ represents the mode-1 unfolding of the second TT-core $\boldsymbol{\mathcal{G}}_2$,  we have
\begin{equation} \boldsymbol{\mathcal{G}}_2 = \boldsymbol{\mathcal{I}}_{3,Q} \times_1 \mathbf{M}_1 \times_2 \mathbf{B}_{\text{sub},D} \times_3 \mathbf{M}_2^{-T}.\end{equation}
By symmetry, applying analogous operations to the right branch yields $\boldsymbol{\mathcal{G}}_3$ and $\boldsymbol{\mathcal{G}}_4$. This concludes the proof. \hfill $\blacksquare$

\section{\textcolor{revisiondarkgreen}{Proof of Theorem 3} \label{app:proof_theorem_3}}

{\color{revisiondarkgreen}
From Theorem 2, the two intermediate TT-cores admit the \textcolor{revisiondarkgreen}{CPD representations}
\begin{equation}
\begin{aligned}
\boldsymbol{\mathcal{G}}_2
&=\llbracket \mathbf{M}_1,\mathbf{B}_{\text{sub},D},\mathbf{M}_2^{-T}\rrbracket,\\
\boldsymbol{\mathcal{G}}_3
&=\llbracket \mathbf{M}_2,\mathbf{B}_{\text{shift},\tau},\mathbf{M}_3^{-T}\rrbracket.
\end{aligned}
\end{equation}
By the essential uniqueness of the \textcolor{revisiondarkgreen}{CPD} of $\boldsymbol{\mathcal{G}}_2$, its recovered factors satisfy
\begin{equation} \label{eq:G2_permutation_ambiguity}
\begin{aligned}
\hat{\mathbf{M}}_1
&=\mathbf{M}_1\boldsymbol{\Pi}_2\mathbf{D}_{2,1},\\
\hat{\mathbf{B}}_{\text{sub},D}
&=\mathbf{B}_{\text{sub},D}\boldsymbol{\Pi}_2\mathbf{D}_{2,2},\\
\hat{\mathbf{M}}_2^{-T}
&=\mathbf{M}_2^{-T}\boldsymbol{\Pi}_2\mathbf{D}_{2,3},
\end{aligned}
\end{equation}
where $\boldsymbol{\Pi}_2$ is a permutation matrix, and $\mathbf{D}_{2,i}$, $i=1,2,3$, are nonsingular diagonal matrices, with $\mathbf{D}_{2,1}\mathbf{D}_{2,2}\mathbf{D}_{2,3}=\mathbf{I}_Q$. Applying the inverse-transpose operation to the last equality in \eqref{eq:G2_permutation_ambiguity}, we obtain
\begin{equation} \label{eq:M2_permutation}
\hat{\mathbf{M}}_2
=\left(\hat{\mathbf{M}}_2^{-T}\right)^{-T}
=\mathbf{M}_2\boldsymbol{\Pi}_2\mathbf{D}_{2,3}^{-1},
\end{equation}
where the last step uses $\boldsymbol{\Pi}_2^{-T}=\boldsymbol{\Pi}_2$, which holds for any permutation matrix.

When recovering $\boldsymbol{\mathcal{G}}_3$, the matrix $\hat{\mathbf{M}}_2$ in \eqref{eq:M2_permutation} is fixed as its first-mode factor. By the essential uniqueness of the \textcolor{revisiondarkgreen}{CPD} of $\boldsymbol{\mathcal{G}}_3$, we have
\begin{equation} \label{eq:G3_M2_permutation}
\hat{\mathbf{M}}_2
=\mathbf{M}_2\boldsymbol{\Pi}_3\mathbf{D}_{3,1}.
\end{equation}
where $\boldsymbol{\Pi}_3$ and $\mathbf{D}_{3,1}$ denote the corresponding permutation matrix and nonsingular diagonal scaling matrix, respectively.
Combining \eqref{eq:M2_permutation} and \eqref{eq:G3_M2_permutation}, we obtain
\begin{equation} \label{eq:permutation_consistency}
\begin{aligned}
\boldsymbol{\Pi}_3^T\boldsymbol{\Pi}_2
&\overset{(a)}{=} \mathbf{D}_{3,1}\mathbf{D}_{2,3}
\overset{(b)}{=} \mathbf{I}_Q.
\end{aligned}
\end{equation}
where $(a)$ follows by canceling the nonsingular matrix $\mathbf{M}_2$ and rearranging the resulting equality, and $(b)$ follows because the only permutation matrix that is also diagonal is $\mathbf{I}_Q$. Consequently, $\boldsymbol{\Pi}_3=\boldsymbol{\Pi}_2\triangleq\boldsymbol{\Pi}$, where $\boldsymbol{\Pi}$ denotes the common permutation matrix.

This equality establishes that the shared transformation matrix $\mathbf{M}_2$ enforces a common column permutation across $\boldsymbol{\mathcal{G}}_2$ and $\boldsymbol{\mathcal{G}}_3$. In particular, the recovered factors satisfy
\begin{equation}
\begin{aligned}
\hat{\mathbf{B}}_{\mathrm{sub},D}
&=\mathbf{B}_{\mathrm{sub},D}\boldsymbol{\Pi}\mathbf{D}_{2,2},\\
\hat{\mathbf{B}}_{\mathrm{shift},\tau}
&=\mathbf{B}_{\mathrm{shift},\tau}\boldsymbol{\Pi}\mathbf{D}_{3,2},\\
\hat{\mathbf{M}}_3^{-T}
&=\mathbf{M}_3^{-T}\boldsymbol{\Pi}\mathbf{D}_{3,3}.
\end{aligned}
\end{equation}

\noindent Finally, by substituting these relations into \eqref{eq:recover_boundary}, we obtain
\begin{equation}
\begin{aligned}
\hat{\mathbf{B}}_{\text{sub},\tau}\hat{\mathbf{\Lambda}}_{|\beta|^2}
&=\mathbf{B}_{\text{sub},\tau}\boldsymbol{\Pi}
\left(\boldsymbol{\Pi}^T\mathbf{\Lambda}_{|\beta|^2}\boldsymbol{\Pi}\right)\mathbf{D}_{2,1},\\
\hat{\mathbf{B}}_{\text{shift},D}
&=\mathbf{B}_{\text{shift},D}\boldsymbol{\Pi}\mathbf{D}_{3,3}.
\end{aligned}
\end{equation}
Since all remaining ambiguities are diagonal, they can be absorbed into $\boldsymbol{\Delta}_1,\ldots,\boldsymbol{\Delta}_4$ in \eqref{eq:common_permutation}. Therefore, all four recovered factors share the same global permutation $\boldsymbol{\Pi}$ and preserve a common target index. This concludes the proof. \hfill $\blacksquare$
}

\makeatletter
\let\revision@bibitem\@bibitem
\def\@bibitem#1{%
  \def\revision@keyA{sun2018thermos}%
  \def\revision@keyB{golub2013matrix}%
  \def\revision@keyC{tensorlab3.0}%
  \def\current@key{#1}%
  \ifx\current@key\revision@keyA
    \color{revisiondarkgreen}%
  \else
    \ifx\current@key\revision@keyB
      \color{revisiondarkgreen}%
    \else
      \ifx\current@key\revision@keyC
        \color{revisiondarkgreen}%
      \else
        \color{black}%
      \fi
    \fi
  \fi
  \revision@bibitem{#1}%
}
\makeatother
\bibliography{ref.bib}

@article{mura2025optimized,
  title={Optimized waveform design for {OFDM}-based {ISAC} systems under limited resource occupancy},
  author={Mura, Silvia and Tagliaferri, Dario and Mizmizi, Marouan and Spagnolini, Umberto and Petropulu, Athina},
  journal={IEEE Trans. Wireless Commun.},
  year={2025},
  publisher={IEEE}
}

@article{li2025sensing,
  title={Sensing-oriented adaptive resource allocation designs for {OFDM-ISAC} systems},
  author={Li, Peishi and Li, Ming and Liu, Rang and Liu, Qian and Swindlehurst, A Lee},
  journal={IEEE Trans. Signal Process.},
  volume={73},
  pages={5121--5135},
  year={2025},
  publisher={IEEE}
}

@article{bicua2016generalized,
  title={Generalized multicarrier radar: {Models} and performance},
  author={Bic{\u{a}}, Marian and Koivunen, Visa},
  journal={IEEE Trans. Signal Process.},
  volume={64},
  number={17},
  pages={4389--4402},
  year={2016},
  publisher={IEEE}
}

@article{zhang2024target,
  title={Target localization in cooperative {ISAC} systems: A scheme based on {5G NR OFDM} signals},
  author={Zhang, Zhenkun and Ren, Hong and Pan, Cunhua and Hong, Sheng and Wang, Dongming and Wang, Jiangzhou and You, Xiaohu},
  journal={IEEE Trans. Commun.},
  year={2024},
  publisher={IEEE}
}

@inproceedings{liu2025low,
  title={Low-Overhead Sensing {RS} Design for Integrated Sensing and Communication},
  author={Liu, Wenjia and Hou, Xiaolin and Liu, Juan and Chen, Lan},
  booktitle={Proc. IEEE Wireless Commun. Netw. Conf. (WCNC)},
  pages={1--6},
  year={2025},
  organization={IEEE}
}

@article{zhang2024ofdm,
  title={{OFDM} reference signal pattern design criteria for integrated communication and sensing},
  author={Zhang, Rui and Tsai, Shawn and Chou, Tzu-Han and Ren, Jiaying and Qu, Wenze and Sun, Oliver},
  journal={IEEE Internet Things J.},
  year={2024},
  publisher={IEEE}
}

@article{tang2021self,
  title={Self-interference-resistant {IEEE} 802.11 ad-based joint communication and automotive radar design},
  author={Tang, Aimin and Li, Songqian and Wang, Xudong},
  journal={IEEE J. Sel. Topics Signal Process.},
  volume={15},
  number={6},
  pages={1484--1499},
  year={2021},
  publisher={IEEE}
}

@article{xie2021performance,
  title={Performance analysis of joint range-velocity estimator with {2D-MUSIC} in {OFDM} radar},
  author={Xie, Rui and Hu, Dengyu and Luo, Kai and Jiang, Tao},
  journal={IEEE Trans. Signal Process.},
  volume={69},
  pages={4787--4800},
  year={2021},
  publisher={IEEE}
}

@article{oseledets2011tensor,
  title={Tensor-train decomposition},
  author={Oseledets, Ivan V},
  journal={SIAM J. Sci. Comput.},
  volume={33},
  number={5},
  pages={2295--2317},
  year={2011},
  publisher={SIAM}
}

@article{biamonte2017tensor,
  title={Tensor networks in a nutshell},
  author={Biamonte, Jacob and Bergholm, Ville},
  journal={arXiv preprint arXiv:1708.00006},
  year={2017}
}

@article{kolda2009tensor,
  title={Tensor decompositions and applications},
  author={Kolda, Tamara G and Bader, Brett W},
  journal={SIAM Rev.},
  volume={51},
  number={3},
  pages={455--500},
  year={2009},
  publisher={SIAM}
}

@article{liu2017hourglass,
  title={Hourglass arrays and other novel {2-D} sparse arrays with reduced mutual coupling},
  author={Liu, Chun-Lin and Vaidyanathan, Palghat P},
  journal={IEEE Trans. Signal Process.},
  volume={65},
  number={13},
  pages={3369--3383},
  year={2017},
  publisher={IEEE}
}

@article{vaidyanathan2010sparse,
  title={Sparse sensing with co-prime samplers and arrays},
  author={Vaidyanathan, Palghat P and Pal, Piya},
  journal={IEEE Trans. Signal Process.},
  volume={59},
  number={2},
  pages={573--586},
  year={2010},
  publisher={IEEE}
}

@article{zhang2025shift,
  title={Shift-Based Symmetric Coprime Planar Arrays With Increased Degrees of Freedom and Reduced Mutual Coupling},
  author={Zhang, Zexiang and Shen, Qing and Liu, Wei and Li, Kangning and Wu, Kejiang},
  journal={IEEE Trans. Veh. Technol.},
  year={2025},
  publisher={IEEE}
}

@article{pal2012nested,
  title={Nested arrays in two dimensions, {{\textcolor{revisiondarkgreen}{Part I}}}: Geometrical considerations},
  author={Pal, Piya and Vaidyanathan, PP},
  journal={IEEE Trans. Signal Process.},
  volume={60},
  number={9},
  pages={4694--4705},
  year={2012},
  publisher={IEEE}
}

@article{gong2025frame,
  title={Frame Structure Design and Task Scheduling in {5G NR} for {ISAC} Systems: {A} Deep Reinforcement Learning Approach},
  author={Gong, Jialong and Gao, Dixiang and Tian, Rui and Liu, Xiqing and Peng, Mugen},
  journal={IEEE Trans. Veh. Technol.},
  year={2025},
  publisher={IEEE}
}

@article{hu2024joint,
  title={Joint range-velocity-azimuth estimation for {OFDM}-based integrated sensing and communication},
  author={Hu, Zelin and Ye, Qibin and Huang, Yixuan and Hu, Su and Yang, Gang},
  journal={IEEE Trans. Wireless Commun.},
  volume={23},
  number={10},
  pages={12933--12948},
  year={2024},
  publisher={IEEE}
}

@article{zheng2023coarray,
  title={Coarray tensor direction-of-arrival estimation},
  author={Zheng, Hang and Zhou, Chengwei and Shi, Zhiguo and Gu, Yujie and Zhang, Yimin D},
  journal={IEEE Trans. Signal Process.},
  volume={71},
  pages={1128--1142},
  year={2023},
  publisher={IEEE}
}

@article{orus2019tensor,
  title={Tensor networks for complex quantum systems},
  author={Or{\'u}s, Rom{\'a}n},
  journal={Nat. Rev. Phys.},
  volume={1},
  number={9},
  pages={538--550},
  year={2019},
  publisher={Nature Publishing Group UK London}
}

@article{evenbly2011tensor,
  title={Tensor network states and geometry},
  author={Evenbly, Glen and Vidal, Guifr{\'e}},
  journal={J. Stat. Phys.},
  volume={145},
  number={4},
  pages={891--918},
  year={2011},
  publisher={Springer}
}

@article{zniyed2019multidimensional,
  title={Multidimensional harmonic retrieval based on {Vandermonde} tensor train},
  author={Zniyed, Yassine and Boyer, Remy and de Almeida, Andr{\'e} LF and Favier, G{\'e}rard},
  journal={Signal Process.},
  volume={163},
  pages={75--86},
  year={2019},
  publisher={Elsevier}
}

@article{gong2023esprit,
  title={An {ESPRIT}-based supervised channel estimation method using tensor train decomposition for {mmWave} {3-D MIMO-OFDM} systems},
  author={Gong, Xiao and Chen, Wei and Sun, Lei and Chen, Jie and Ai, Bo},
  journal={IEEE Trans. Signal Process.},
  volume={71},
  pages={555--570},
  year={2023},
  publisher={IEEE}
}

@techreport{ITURRECP525, 
	author = {ITU}, 
	institution = {{International Telecommunication Union (ITU)}}, 
	month = aug, 
	day={11},
	note = {}, 
	number = {{P.525-4}}, 
	title = {Calculation of free-space attenuation }, 
	type = {Recommendation (R)}, 
	url = {https://www.itu.int/rec/R-REC-P.525}, 
	year = {2019}
}

@techreport{3GPP138104, 
	author = {3GPP}, 
	institution = {{3rd Generation Partnership Project (3GPP)}}, 
	month = oct, 
	day={13},
	note = {Version 19.2.0}, 
	number = {38.104}, 
	title = {{5G}; {NR}; Base Station ({BS}) radio transmission and reception}, 
	type = {Technical Specification (TS)}, 
	url = {https://www.etsi.org/deliver/etsi_ts/138100_138199/138104/}, 
	year = {2025}
}

@article{tang2025cooperative,
  title={Cooperative {ISAC}-empowered low-altitude economy},
  author={Tang, Jun and Yu, Yiming and Pan, Cunhua and Ren, Hong and Wang, Dongming and Wang, Jiangzhou and You, Xiaohu},
  journal={IEEE Trans. Wireless Commun.},
  year={2025},
  publisher={IEEE}
}

@article{kumari2019adaptive,
  title={Adaptive virtual waveform design for millimeter-wave joint communication--radar},
  author={Kumari, Preeti and Vorobyov, Sergiy A and Heath, Robert W},
  journal={IEEE Trans. Signal Process.},
  volume={68},
  pages={715--730},
  year={2019},
  publisher={IEEE}
}

@inproceedings{wang2025framework,
  title={A framework for {{ISAC}} systems using non-uniformly distributed sensing resources},
  author={Wang, B. and Pan, C. and Ren, H. and Yu, Z.},
  booktitle={Proc. 17th Int. Conf. Wireless Commun. Signal Process. (WCSP)},
  year={2025},
  pages={1--6}
}

@article{ma2021multi,
  title={Multi-source off-grid {DOA} estimation with single snapshot using non-uniform linear arrays},
  author={Ma, Yanan and Cao, Xianbin and Wang, Xiangrong and Greco, Maria S and Gini, Fulvio},
  journal={Signal Process.},
  volume={189},
  pages={108238},
  year={2021},
  publisher={Elsevier}
}

@article{zhou2017low,
  title={Low-rank tensor decomposition-aided channel estimation for millimeter wave {MIMO-OFDM} systems},
  author={Zhou, Zhou and Fang, Jun and Yang, Linxiao and Li, Hongbin and Chen, Zhi and Blum, Rick S},
  journal={IEEE J. Sel. Areas Commun.},
  volume={35},
  number={7},
  pages={1524--1538},
  year={2017},
  publisher={IEEE}
}

@misc{tensorlab3.0,
  title={{Tensorlab 3.0}},
  author={Vervliet, N. and Debals, O. and Sorber, L. and Van Barel, M. and De Lathauwer, L.},
  howpublished={Online},
  month=mar,
  note={Available: \url{https://www.tensorlab.net}},
  year={2016}
}

@article{ren2018extensions,
  title={Extensions of open box array with reduced mutual coupling},
  author={Ren, Shiwei and Li, Xiangnan and Luo, Xue and Wang, Weijiang},
  journal={IEEE Sensors J.},
  volume={18},
  number={13},
  pages={5475--5484},
  year={2018},
  publisher={IEEE}
}

@article{sun2018thermos,
  title={Thermos array: Two-dimensional sparse array with reduced mutual coupling},
  author={Sun, Leilei and Yang, Min and Chen, Baixiao},
  journal={Int. J. Antennas Propag.},
  volume={2018},
  pages={1--9},
  year={2018},
  publisher={Wiley}
}

@article{liu2022integrated,
  title={Integrated sensing and communications: {Toward} dual-functional wireless networks for {6G} and beyond},
  author={Liu, Fan and Cui, Yuanhao and Masouros, Christos and Xu, Jie and Han, Tony Xiao and Eldar, Yonina C and Buzzi, Stefano},
  journal={IEEE J. Sel. Areas Commun.},
  volume={40},
  number={6},
  pages={1728--1767},
  year={2022},
  publisher={IEEE}
}

@article{cui2021integrating,
  title={Integrating sensing and communications for ubiquitous {IoT}: {Applications}, trends, and challenges},
  author={Cui, Yuanhao and Liu, Fan and Jing, Xiaojun and Mu, Junsheng},
  journal={IEEE Netw.},
  volume={35},
  number={5},
  pages={158--167},
  year={2021},
  publisher={IEEE}
}

@article{liu2020joint,
  title={Joint radar and communication design: Applications, state-of-the-art, and the road ahead},
  author={Liu, Fan and Masouros, Christos and Petropulu, Athina P and Griffiths, Hugh and Hanzo, Lajos},
  journal={IEEE Trans. Commun.},
  volume={68},
  number={6},
  pages={3834--3862},
  year={2020},
  publisher={IEEE}
}

@article{lu2024integrated,
  title={Integrated sensing and communications: Recent advances and ten open challenges},
  author={Lu, Shihang and Liu, Fan and Li, Yunxin and Zhang, Kecheng and Huang, Hongjia and Zou, Jiaqi and Li, Xinyu and Dong, Yuxiang and Dong, Fuwang and Zhu, Jia and others},
  journal={IEEE Internet Things J.},
  volume={11},
  number={11},
  pages={19094--19120},
  year={2024},
  publisher={IEEE}
}

@techreport{3GPP13810101, 
	author = {3GPP}, 
	institution = {{3rd Generation Partnership Project (3GPP)}}, 
	month = aug, 
	day={25},
	note = {Version 18.10.0}, 
	number = {38.101}, 
	title = {{5G}; {NR}; User Equipment ({UE}) radio transmission and reception; Part 1: Range 1 Stand alone}, 
	type = {Technical Specification (TS)}, 
	url = {	https://www.etsi.org/deliver/etsi_ts/138100_138199/13810101/}, 
	year = {2025}
}

@article{ni2021uplink,
  title={Uplink sensing in perceptive mobile networks with asynchronous transceivers},
  author={Ni, Zhitong and Zhang, J Andrew and Huang, Xiaojing and Yang, Kai and Yuan, Jinhong},
  journal={IEEE Trans. Signal Process.},
  volume={69},
  pages={1287--1300},
  year={2021},
  publisher={IEEE}
}

@article{wu2024sensing,
  title={Sensing in bistatic {ISAC} systems with clock asynchronism: A signal processing perspective},
  author={Wu, Kai and Pegoraro, Jacopo and Meneghello, Francesca and Zhang, J Andrew and Lacruz, Jesus O and Widmer, Joerg and Restuccia, Francesco and Rossi, Michele and Huang, Xiaojing and Zhang, Daqing and others},
  journal={IEEE Signal Process. Mag.},
  volume={41},
  number={5},
  pages={31--43},
  year={2024},
  publisher={IEEE}
}

@article{li2024mimo,
  title={{MIMO-OFDM ISAC} waveform design for range-{Doppler} sidelobe suppression},
  author={Li, Peishi and Li, Ming and Liu, Rang and Liu, Qian and Swindlehurst, A Lee},
  journal={IEEE Trans. Wireless Commun.},
  volume={24},
  number={2},
  pages={1001--1015},
  year={2024},
  publisher={IEEE}
}

@article{zhu2009chunk,
  title={Chunk-based resource allocation in {OFDMA} systems-{Part I}: chunk allocation},
  author={Zhu, Huiling and Wang, Jiangzhou},
  journal={IEEE Trans. Commun.},
  volume={57},
  number={9},
  pages={2734--2744},
  year={2009},
  publisher={IEEE}
}

@book{golub2013matrix,
  author={Golub, Gene H. and Van Loan, Charles F.},
  title={Matrix Computations},
  edition={4},
  publisher={Johns Hopkins University Press},
  address={Baltimore, MD, USA},
  year={2013}
}
\bibliographystyle{ieeetr}

\end{document}